\documentclass[%
 reprint,
 amsmath,amssymb,
 aps,
floatfix,
]{revtex4-2}

\usepackage[title]{appendix}
\usepackage{amsmath}   % <--- 
 \usepackage{amsfonts}
\usepackage{amssymb}
\usepackage{multirow}
\usepackage{booktabs}
\usepackage{array}
\usepackage{graphicx}% Include figure files
\usepackage{float}
\usepackage[export]{adjustbox} % Allows cropping and positioning
\usepackage{subcaption}
\usepackage{subfiles}
\usepackage{braket}
\usepackage{svg}
\usepackage{placeins} 
\usepackage{tikz}
    \usetikzlibrary{quantikz}
\usepackage{dcolumn}% Align table columns on decimal point
\usepackage{bm}
\usepackage{algorithm}
\usepackage{algpseudocode}
\usepackage{dcolumn}% Align table columns on decimal point
\usepackage{bm}% bold math
\usepackage{xcolor}
\begin{document}

\preprint{APS/123-QED}

\title{Quantum lattice Boltzmann method for several time steps: A local Carleman linearization algorithm }
\author{Antonio David Bastida Zamora}
\altaffiliation{Also at Aix-Marseille University, France}
\affiliation{Quanscient Oy, Finland}

\author{Ljubomir Budinski}
\altaffiliation{Also at Faculty of Technical Sciences, University of Novi Sad, Serbia}
\affiliation{Quanscient Oy, Finland}

\author{Pierre Sagaut}
\altaffiliation{Also at Quanscient Oy, Finland}
\affiliation{Aix Marseille Univ, Centrale Med, CNRS, M2P2 Laboratory, 13013 Marseille, France}

\author{Valtteri Lahtinen}
\altaffiliation{Also at School of Engineering Science, Lappeenranta–Lahti University of Technology, Finland}
\affiliation{Quanscient Oy, Finland}

\date{\today}
\begin{abstract}
This article presents a novel encoding for quantum Lattice Boltzmann method algorithm using Carleman linearization. In contrast to previous articles \cite{Sanavio2024LatticeBC,sanavio2025carleman}, the encoding used allows for local collision rules while keeping a higher probability to obtain the right result, which is of the order of $10^{-2}$. The algorithm scales as $O(log_2^2(N)+Q^3)$ each time step with $N$ the number of lattice sites of the 2D lattice and $Q$ the number of channels with a constant number of qubits when using dynamical circuits.

\end{abstract}

\maketitle

\section{Introduction}

Lattice Boltzmann method (LBM), originally developed to solve Navier-Stokes equations, has been successfully applied to many other fields including electromagnetics, biology, and material science \cite{PhysRevE.99.033301,su13147968,app12094627,escande2020lattice,muller2023dynamic}. Despite the method's success and the existence of rare landmark achievements in terms of Extreme Computing such as \cite{xu2025towards} in which simulations with up to $2\cdot10^{12}$ lattices using $1.55\cdot10^8$ heterogeneous many-core CPUs or \cite{Falcucci2021} with about 50 billion
lattice sites and ledding to substantial new physics, a larger scale was needed. Simulations with the number of lattice sites required to handle full-scale engineering applications with a completely detailed geometry  are still out of reach for daily use due to the high computational cost associated with them. 
In computational fluid dynamics (CFD), a high number of lattice sites allows for the obtaining of a high Reynolds number, which is necessary for many impactful industrial applications such as the aerospace and automotive sectors. To solve this issue, researchers began to explore possible paths to increase the number of lattice sites without losing accuracy, finding the usage of quantum computers as an interesting alternative. 

The first quantum algorithms for LBM (that tried to replicate results from classical LBM with quantum circuits) were applied to simple linear cases such as advection-diffusion \cite{Budinski_2021}. In this article, an algorithmic advantage in the complexity of quantum LBM compared to the classical methods was reported for the first time. The algorithm comprises collision, propagation and calculation of macroscopics, scaling as $O(log^2(N))$. However, the algorithm was limited to a single time step, and the state preparation (encoding of classical information into the quantum system) and the measurement (extraction of quantum information to the classical system) scaled linearly in depth $O(N)$, which did not contribute to the advantage sought. Soon after, a quantum algorithm for Navier-Stokes equations using a novel stream-vorticity encoding was proposed, building the quantum circuit to obtain the same results as LBM for nonlinear cases. Despite that, the limitations previously introduced were not tackled. In recent years, QLBM has grown in popularity among researchers, with an increasing number of articles on the subject. For instance, we have seen new methods to extract results \cite{schalkers2024momentum}, projects for open-source frameworks \cite{GEORGESCU2025109699}, 3D simulations under more complex cases \cite{xiao2025quantum} and time-stepping for linear cases using dynamical circuits and linear combination of unitaries (LCU) \cite{wawrzyniak2025dynamic}. Nonetheless, these methods did not achieve an optimal algorithm for QLBM with several time steps.

Inspired by the limitations of other nonlinear QLBM simulations, Sanavio \textit{et al.} began to consider the necessity of creating a general linear QLBM algorithm, able to give similar results but only using a system of linear equations. This was achieved by using Carleman linearization (CL), an approach used in nonlinear partial differential equations to simplify them into linear systems \cite{Sanavio2024LatticeBC}. However, this first attempt to create a quantum circuit to handle LBM for nonlinear cases at second order CL had several issues. 

First, the circuit depth scales as $O(N^2 Q^4)$ with $Q$ the number of channels and $N$ the number of lattice sites of the 2D lattice. The reason for this is the nonlocality of the collision operator. In this context, nonlocality is defined by the collision operator's depth scaling, which is unstructured relative to the number of lattice sites $N$ and does not take advantage of quantum superposition principle. This means, each quantum register for lattice sites is modified individually and not all at once. Notice that while in LBM, each lattice site follows the same collision operator, when using Carleman linearization, different orders must be computed differently. Additionally, the lattice site encoded at different Carleman orders is different, needing a nontrivial operator for each lattice site, to prepare the correct encoding. Second, the probability of measuring the correct result at every time step is moderately low, around $10^{-2}$, depending on the viscosity and initial state of the system. To solve the depth problem, the authors proposed a novel Navier-Stokes approach using Carleman Method and a Carleman-Grad approach \cite{sanavio2024carleman,sanavio2024three}. The first method, while reducing the depth, introduced stability problems. These problems impact the convergence of the simulation. The Carleman-Grad approach introduces approximations that significantly decrease the depth but includes lower accuracy, which in many cases results in the need for higher orders in the linearization. Recently, a new approach based on matrix access oracles was proposed \cite{sanavio2025carleman}, erasing the problem of nonlocality but decreasing the probability to measure the right outcome to $10^{-5}$ for a single time-step, which makes practical applications suboptimal. Despite the present challenges, recent studies show that a modest quantum advantage using quantum Carleman linearization for 2D can be obtained \cite{Jennings2025}. The authors review carefully every step of the quantum Carleman algorithm applied to LBM, concluding that a quantum advantage cannot be obtained for high Reynolds numbers and general problems. Despite that, the authors argue that for low Reynolds numbers and certain problems where Carleman provides sufficient convergence, an advantage could be achieved. This new evidence reinforces the need for more efficient quantum encodings for Carleman linearization.

In this paper, we propose a new quantum encoding for Carleman QLBM that preserves locality and increases the probability of measuring the right outcome to the order of $10^{-2}$. First, in section~II, we introduce LBM and Carleman methods. Section~III introduces QLBM using Carleman linearization and the new encoding to obtain locality and analyzes its computational complexity. Finally, in section~IV, we share the results and validation, before concluding in secction~V.

\section{Carleman linearization for Lattice Boltzmann Method}
\label{sec:car_class}
The Lattice Boltzmann Method (LBM) was initially derived from Lattice Gas Automata (LGA) as a more efficient and accurate alternative to simulate the Navier-Stokes equations for fluid dynamics. An introduction to LGA can be found in \cite{Rivet2001}. Lattice gas automata is a cellular automata model in which all interactions between particles are local and simultaneous and obey mass and momentum conservation. The model can be considered a direct numerical method, where each fictitious particle in a fluid is simulated. However, the high number of particles, collisions, and noise in the model makes it expensive. LBM appeared as a more optimal and accurate alternative to directly simulate the Boltzmann equation at a mesoscopic level \cite{BENZI1992145,Higuera_1989}. It is obtained when taking the ensemble average of particles in LGA. In this new method, density probability distributions of particles are calculated instead of the particles themselves. Given this physical interpretation, we can introduce LBM as an algorithm based on continuous density functions $f_i$ associated with each velocity $v_i$. Each density function is defined in a lattice site $x$ and a discrete time step $t$. Therefore, the distribution function $f$ can be expressed as a vector with components $f_i(x,t)$.

\begin{figure}
    \centering
    \includegraphics[width=0.15\textwidth]{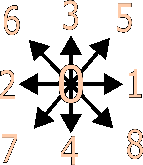}

    \caption{Scheme of channels for D2Q9. Each number represents the index used to map each channel.} 
    \label{fig:d2q9}
\end{figure}

There are many different variants of the LBM proposed in the bibliography. To simplify the analysis and be specific, we will make use of the Bhatnagar-Gross-Krook approximation (BGK) with a D2Q9 scheme following Fig~\ref{fig:d2q9}, which reduces the evolution of the distribution functions to 
\begin{equation}
    f_i(x+v_i,t+1)=f_i(x,t)+\Omega_i
\end{equation}
$\Omega_i$ being the collision operator, which takes the form of 
\begin{equation}
    \Omega_i=\frac{1}{\tau} \left(f_i^{eq}(x,t)-f_i(x,t)\right)
\end{equation}
with $f_i^{eq}$ the local equilibrium distribution function that each $f_i$ will relax to and $\tau$ the relaxation time. If we write the equilibrium distribution functions using a Taylor expansion at second order, we obtain
\begin{equation}
f_i^{eq}(x,t)=w_i \rho(x,t) \left(1+\frac{u c_i}{c_s^2}+\frac{(uc_i)^2}{2c_s^4}-\frac{u u}{2c_s^2}\right)
\end{equation}
where $c_i$ is the direction of the propagation for the channel $i$, being 1 for positive direction and -1 for negative direction, $c_s$ the lattice speed of sound, in our case equal to $\frac{1}{\sqrt{3}}$, $w_i$ are the weights associated to each channel (calculated to obtain the right symmetry relations) and $u(x,t)$ the local velocity defined as
\begin{equation}
    u(x,t)=\frac{\sum\limits_{i=0}^{Q-1}f_i(x,t)c_i}{\sum\limits_{i=0}^{Q-1}f_i(x,t)}
\end{equation}
The Carleman linearization (CL) is a method to write a nonlinear dynamical system into an infinite linear system of equations. In this regard, we can write products of local distribution functions $f_i$ as $g_{ij}(x_1,x_2)=f_i(x_1)f_j(x_2)$, $h_{ijk}(x_1,x_2,x_3)=f_i(x_1)f_j(x_2)f_k(x_3)$... As $f_i=f_i(f,g,h,...)$, a dynamical variable of order $n$ will only depend on dynamical variables of the same or higher order. While an infinite number of terms will give the correct result, we can also approximate this system by including a bound on this series. For example, if we only use $f$ and $g$, we will obtain a system of two linear equations and use a second-order CL. In practice, CL can be used by finding a specific order that provides the desired accuracy. In some cases, where the dependence of the highest order of the series is trivial, a closure approximation is done. This means that the highest order is approximated depending on smaller orders. This approximation can be derived explicitly from the properties of the system to study. In this article, we focus on its application to LBM. For a more comprehensive and formal treatment of the method, we refer the reader to \cite{kowalski1991nonlinear}.

The first attempt to use a Carleman linearization for a quantum implementation of Lattice Boltzmann Method (LBM) was \cite{Sanavio2024LatticeBC} by Succi and Sanavio. In this paper, the authors introduce the key operators to encode the local equilibrium distribution functions. Specifically, for a weakly-compressible flow
\begin{equation}
    f_i^{eq}=L_{ij}f_j+Q_{ijk} f_j f_k + T_{ijkl} f_j f_k f_l
\end{equation}
 with a linear, quadratic and cubic contributions, where the density is assumed to be nearly constant, with a value $\rho\approx1$, following the weakly compressible limit. This assumption will add a low computational cost, given that we will focus on flows with low Mach number and that $\delta \rho =O(\frac{u^2}{c_s^2})$. The operators are
\begin{equation}
\begin{aligned}
L_{ij} &= w_i\left(1+\frac{c_i \cdot c_j}{c_s^2}\right) \\
Q_{ijk} &= \frac{w_i}{c_s^4}(c_i \cdot c_j c_i \cdot c_k- c_s^2 c_j \cdot c_k) \\
T_{ijkl} &= -\frac{1}{2}Q_{ijk} \quad \forall l
\end{aligned}
\end{equation}
and using the LBM collision equation
\begin{equation}
f_i(x+\delta x,t+\delta t)=f_i(x,t)+\omega (f_i^{eq}(x,t)-f_i(x,t))
\end{equation}
we can obtain the collision equation
\begin{equation}
    f_i=A_{ij}f_j+B_{ijk} f_j f_k + C_{ijkl} f_j f_k f_l
\end{equation}
 with 
\begin{equation}
\left\{
\begin{array}{l}
A_{ij} = (1-\omega) \delta_{ij} + \omega L_{ij}, \\
B_{ijk} = \omega Q_{ijk}, \\
C_{ijkl} = -\frac{\omega}{2} Q_{ijk} \quad \forall l
\end{array}
\right.
\end{equation}
The Carleman procedure now uses nonlinear terms of the local distribution function $f_i$ as a dynamic variable. Therefore, solving a system of equations whose order corresponds to the truncation. Without loss of generality, the authors focus on a second-order CL, obtaining
\begin{equation}
\begin{aligned}
f_i(x_1) &=A_{ij} f_j(x_1) + B_{ijk} g_{jk} (x_1,x_1)\\
g_{i,j}(x_1,x_2)&=A_{ik}A_{jl}g_{kl}(x_1,x_2)
\end{aligned}
\end{equation}
The cited article introduces accurate results using the Carleman method for classical LBM for small systems with moderate Reynolds numbers. These results constitute evidence of the method's validity. However, this technique uses many variables growing as $O(N^2)$, with $N$ the number of lattice sites in $f$, making the problem intractable for classical computers. Quantum computers could solve this issue by using the superposition principle, with the number of qubits scaling as $O(log^2(N))$.

\section{Quantum algorithm for LBM using Carleman linearization}
\label{sec:quantum}
In the first article introducing QLBM with Carleman \cite{Sanavio2024LatticeBC}, the authors introduce a simple embedding that optimally encodes all the Carleman variables. This embedding will be manipulated using unitary operators via linear algebra. For this purpose they use 5 registers to encode each variable: $\ket{x_1}_{p_1}, \ket{x_2}_{p_2},\ket{i}_{c_1}, \ket{j}_{c_2},\ket{0/1}_\tau$, encoding the first lattice, second lattice, first channel, second channel and the Carleman order respectively. Notice that the symbol in the ket is the variable, while the subscript refers to the register notation. For a second-order Carleman linearization, we have
\begin{equation}
\begin{aligned}
    \ket{f} &= \sum_{x=0}^{N-1} \sum_{i=0}^{Q-1} \alpha_{x_i} \ket{x}_{p_1}\ket{0}_{p_2} \ket{i}_{c_1}\ket{0}_{c_2}\ket{0}_\tau\\
    \ket{g} &= \sum_{x_1,x_2=0}^{N-1}\sum_{i,j=0}^{Q-1}\beta_{{x_{1}}_i {x_{2}}_j}\ket{x_1}_{p_1}\ket{x_2}_{p_2}\ket{i}_{c_1}\ket{j}_{c_2}\ket{1}_\tau 
\end{aligned}
\end{equation}
where $N$ is the number of lattice sites in each lattice register. After the initial state preparation, the system is collided. A procedure based on a linear combination of unitaries (LCU) is used to realise the collision. This is because the operator presented in Sec~\ref{sec:car_class} is not unitary; this technique allows us to make it unitary by embedding it into a larger system using an ancilla, obtaining the correct result with a certain probability. The probability of measuring the correct output will depend on the exact implementation of this procedure and the initial state of the simulation. In summary, the global operator can be decomposed as
\begin{equation}
    C=U_a+\gamma U_b
\end{equation}
with $C$ written in terms of projections for each register as
\begin{equation}
\label{eq:operator}
\begin{aligned}
&C = \ket{x_1}\bra{x_1}_{p_1} \otimes\ket{0}\bra{0}_{p_2} \otimes \sum_{ij} A_{ij} \ket{i}\bra{j}_{c_1} \otimes \ket{0}\bra{0}_{c_2} \otimes \ket{0}\bra{0}_\tau \\
&+ \ket{x_1}\bra{x_1}_{p_1} \otimes \ket{0}\bra{x_2}_{p_2} \otimes \sum_{ijk} B_{ijk} \ket{i}\bra{j}_{c_1} \otimes \ket{0}\bra{k}_{c_2} \otimes \ket{0}\bra{1}_\tau \\
&+ \ket{x_1}\bra{x_1}_{p_1} \otimes \ket{x_2}\bra{x_2}_{p_2} \otimes \sum_{ijkl} A_{ik} \ket{i}\bra{k}_{c_1} \otimes  A_{jl} \ket{j}\bra{l}_{c_2} \otimes \ket{1}\bra{1}_\tau
\end{aligned}
\end{equation}
As we notice, every term depends in a nontrivial way on $p_2$. This means that we need to apply multiple CNOT  gates to the register $p_2$, which translates to a rapid growth of the depth with the number of lattice sites. Because of this, the algorithm scales as $O(N^4Q^4)$ in two-qubit gates. This is because the operator is not written in block-diagonal form.

\subsection{Local quantum Carleman algorithm}
This section will introduce a novel encoding to make the collision \eqref{eq:operator} local. As we have seen, to make the collision operator $C$ local, we must make it trivially dependent on the register $p_2$. Every contribution in the register $p_2$ must be $I_{p_2}$. Which translates to 
\begin{equation}
\begin{aligned}
C &= I_{p_1} \otimes I_{p_2} \otimes \sum_{ij} A_{ij} \ket{i}\bra{j}_{c_1} \otimes \ket{0}\bra{0}_{c_2} \otimes \ket{0}\bra{0}_\tau \\
&\quad + I_{p_1} \otimes I_{p_2} \otimes \sum_{ijk} B_{ijk} \ket{i}\bra{j}_{c_1} \otimes \ket{0}\bra{k}_{c_2} \otimes \ket{0}\bra{1}_\tau \\
&\quad + I_{p_1} \otimes I_{p_2} \otimes \sum_{ik} A_{ik} \ket{i}\bra{k}_{c_1} \otimes \sum_{jl} A_{jl} \ket{j}\bra{l}_{c_2} \otimes \ket{1}\bra{1}_\tau
\end{aligned}
\end{equation}
To accomplish this, we need to make the term $f(t)$ that contributes to $f(t+1)$ and the term $g(t)$ that contributes to $g(t+1)$ independent of the register $p_2$. This means that $p_2$ should be constant and not be changed during the collision, similar to $p_1$ in the operator. One may think we could do this without any issue, since the second register for $f$ does not play any role in the propagation or collision. The reason why this is problematic, is that if we do not apply the projection $\ket{0}\bra{x}_{p_2}$, some terms for $f$ will be $\ket{0}_{p_2}\ket{0}_\tau$ and others $\ket{x}_{p_2}\ket{0}_\tau$, needing to apply the collision operator in all possible states for $p_2$, which makes the two-qubit gates to scale as $O(N^4)$. Fixing this requires the following encoding of the initial state
\begin{equation}
\begin{aligned}
\ket{\psi} =\; & \sum_{x_1=1}^N \sum_{i=1}^Q \alpha_{x_i}\ket{x}_{p_1} \ket{0}_{p_2} \ket{i}_{c_1} \ket{0}_{c_2} \ket{0}_{\tau} \\
& + \sum_{x_1=1}^N \sum_{i=1}^Q \sum_{j=1}^Q \beta_{{x_{1}}_i 0_{j}} \ket{x_1}_{p_1} \ket{0}_{p_2} \ket{i}_{c_1} \ket{j}_{c_2} \ket{1}_{\tau} \\
& + \sum_{x_1,x_2=1}^N \sum_{i=1}^Q \sum_{j=1}^Q  \beta_{{x_{1}}_i {x_{2}}_j}\ket{x_1}_{p_1} \ket{x_2}_{p_2} \ket{i}_{c_1} \ket{j}_{c_2} \ket{1}_{\tau}.
\end{aligned}
\label{eq:state_local}
\end{equation}
Using \eqref{eq:state_local}, we can apply $I_{p_2}$ as the quantum states with register $\ket{0}_{\tau}$ will always have $\ket{0}_{p_2}$. The diagonal terms, corresponding to $g(x_1,x_1)$ are stored in $\ket{0}_{p_2}\ket{1}_{\tau}$. After changing the register $p_2$ for diagonal terms, we arrange the rest in other positions in the quantum state, conserving their neighbourhood. In the case of 1D, this is trivially done by shifting all the states where $g(x_1,x_2)\rightarrow \ket{x_1}_{p_1}\ket{x_2-x_1}_{p_2}\ket{1}_{\tau}$, which preserves periodic boundary conditions. For the case of 2D, we can use the same encoding as 1D for each component. Therefore using $x_1=(x_{11},x_{12})$ and $x_2=(x_{21},x_{22})$, we have $g(x_1,x_2)\rightarrow \ket{x_{11},x_{12}}_{p_1}\ket{(x_{21}-x_{11})\bmod L_x,(x_{22}-x_{12})\bmod L_y}_{p_2}\ket{1}_{\tau}$, with $L_x$ and $L_y$ the number of lattice sites in each dimension of the 2D lattice such that $N=L_xL_y$. However, for the 2D case, the coupling between the $x$ and $y$ directions for the first and second lattice registers $p_1$ and $p_2$ modifies how the propagation is implemented in the algorithm. After the usual shift in the first and second registers, an additional shift is applied on $p_2$ conditioned on $p_1$ and $\ket{1}_\tau$ is realised. The total shift operator $S_T$ for $g$ takes the form of
\begin{equation}
    S_T\ket{x_1}_{p_1}\ket{x_2}_{p_2}\ket{1}_{\tau}\rightarrow \ket{x_1+u_1}_{p_1}\ket{x_2+u_2-u_1}_{p_2}\ket{1}_{\tau}.
\end{equation}
In the appendix~\ref{sec:apx}, we show a proof regarding the permutation of terms in the encoding described and how the operator modifies the quantum state vector.

The algorithm's pseudocode can be seen in \ref{pseudocode}, where we used the notation $\ket{x_1}_{p_1} \ket{x_2}_{p_2} \ket{i}_{c_1} \ket{j}_{c_2} \ket{0/1}_{\tau}=:\ket{c_v(x_1,x_2,i,j,0/1)}$ for simplification. 

\begin{algorithm}[H]
\caption{Quantum LBM Evolution}
\begin{algorithmic}[1]
\State \textbf{Initialization} \Comment{Scaling: $\mathcal{O}(NQ)$}
\For{$x_{11} = 1$ to $L_x$}
    \For{$x_{12} = 1$ to $L_y$}
        \State $x_1 \gets x_{12} + x_{11} L_y$
        \State $\ket{c_v(x_1, 0, i, 0, 0)} \gets f_{i}(x_1)$
    \EndFor
\EndFor
\State $\ket{c_v(x_1, x_2, i, j, 1)} \gets g_{i,j}(x_1, x_2)$ \Comment{Using $CR_y$ and tensor product of $f$}

\State \textbf{Permutation of terms} \Comment{Scaling: $\mathcal{O}(\log_2^3(N))$}
\State $(x_{11},x_{12}) \gets x_1$
\State $(x_{21},x_{22}) \gets x_2$

\For{$k = 1$ to $\log^2(L_x)$}
    \If{$bin(x_{11})[k] == 1$}
        \State $\ket{c_v(x_1, (x_{21}-2^k,x_{22}), i, j, 1)} \gets \ket{c_v(x_1, (x_{21},x_{22}), i, j, 1)}$
    \EndIf
\EndFor
\For{$k = 1$ to $\log^2(L_y)$}
    \If{$bin(x_{12})[k] == 1$}
        \State $\ket{c_v(x_1, (x_{21},y_{22}-2^k), i, j, 1)} \gets \ket{c_v(x_1, (x_{21},x_{22}), i, j, 1)}$
    \EndIf
\EndFor

\For{$t = 1$ to $T$}
    \State \textbf{Collision} \Comment{Scaling: $\mathcal{O}(Q^4 + \log_2^2(N))$}
    \State $\ket{c_v(x_1, 0, i, 0, 0)} \gets A \ket{c_v(x_1, 0, i, 0, 0)}$
    \State $\ket{c_v(x_1, 0, i, 0, 0)} \gets \ket{c_v(x_1, 0, i, 0, 0)} + B \ket{c_v(x_1, 0, i, j, 1)}$
    \State $\ket{c_v(x_1, x_2, i, j, 1)} \gets A \otimes A \ket{c_v(x_1, x_2, i, j, 1)}$

    \State \textbf{Propagation} \Comment{Scaling: $\mathcal{O}(\log_2^2(N))$}
    \State $\ket{c_v(x_1 + u_1, 0, i, 0, 0)} \gets \ket{c_v(x_1, 0, i, 0, 0)}$
    \State $\ket{c_v(x_1 + u_1, x_2 + u_2 - u_1, i, j, 1)} \gets  \ket{c_v(x_1, x_2, i, j, 1)}$
\EndFor

\State \textbf{Measurement} \Comment{Scaling: $\mathcal{O}(NQ)$}
\end{algorithmic}
\label{pseudocode}
\end{algorithm}

\subsection{Computational steps and complexity}
\begin{figure}
	\centering
	\resizebox{0.5\textwidth}{!}{\input{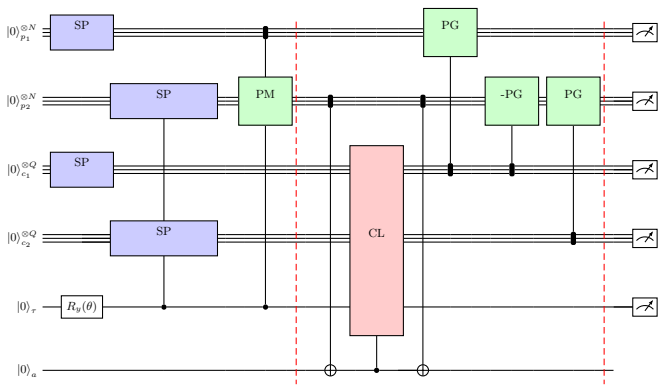}}
    \caption{Quantum circuit scheme for QLBM using Carleman. The circuit is composed of the state preparation (SP), collision (CL), propagation (PG) and permutation (PM) operators. The dashed lines encompass the section of the circuit that is repeated $T$ time steps. Notice that the black dots in the quantum gates represent several conditions over the given register.}
    \label{fig:scheme_algorithm}
\end{figure}

In this section, we will review the different steps of the algorithm and their computational complexity. We will enumerate each section of the algorithm and discuss its scaling. Figure~\ref{fig:scheme_algorithm} shows an overview of the algorithm.

\begin{enumerate}
    \item \textbf{State preparation:} To encode the initial state of the lattice, we can use any state preparation algorithm. The bibliography in this regard is extensive, and the problem of encoding data into a quantum circuit remains one of the most challenging and expensive subroutines of quantum algorithms in general \cite{mottonen2005transformation}. Recent works have focused on using ancillas or approximations to decrease this depth to a logarithmic scaling  \cite{araujo2021divide,PhysRevLett.129.230504}. Other works focus on specific functions to initialize, where logarithmic scaling is possible \cite{PhysRevA.109.042401}. In general, for our purpose, where the initial state can be very general and quantum circuits need to be easily constructed, we assume here a linear scaling with the lattice sites $O(NQ)$. However, this will be problem-specific and strategies based on increasing the number of time-steps with simpler initial conditions may be applicable. As part of the initialization step, we also include the encoding of boundary conditions, which in case they exist, will be constant during the simulation and therefore only done once. These boundaries can be encoded with an extra ancilla qubit. Without loss of generality, we suppose that the collision and propagation are homogeneous and therefore we include no boundaries. Specifically, the state preparation is based on the encoding of $f$ in the first lattice and channel register $p_1$ and $c_1$.
    \item \textbf{Permutation of terms and encoding of $g$:}  Once we have encoded our initial distribution $f$, we need to encode $g$. For that we begin with the initialization of the distribution $f$ in the second lattice and channel register $p_2$ and $c_2$, respectively. The state preparation is conditioned on the $\tau$ register. To encode the relative amplitude of $g$ and $f$ in the quantum state, we make use of a controlled $R_y$ gate, with angle $\alpha=2 \mathrm{arcsin}(\sqrt{p})$ with $p=\frac{\sum\limits_{x,i}f_i(x)^2}{1+\sum\limits_{x,i}f_i(x)^2}$. Notice that this step scales as $O(NQ)$ and therefore, does not add any classical overhead. After this, the need to execute the permutation of terms in the quantum state. We apply a shift routine to the second lattice register $p_2$ controlled by the first lattice register $p_1$ for the terms where $\tau=1$. If the $n^{th}$ qubit is 1, we subtract $2^n$ from the second lattice register. Therefore, we need $log(N)$ number of shifts, with each of them scaling as $O(log^2(N))$ according to the state-of-the-art \cite{Budinski2023ParallelQuantumShift}, obtaining a final complexity of $O(log^3(N))$. The quantum circuit for this step can be seen in Fig~\ref{fig:scheme_permutation}. Notice that the figure for the registers $p_1$ and $p_2$ corresponds to the $x$-dimension. The same is done for the $y$-dimension.

    \begin{figure}
	\centering
	\resizebox{0.5\textwidth}{!}{\input{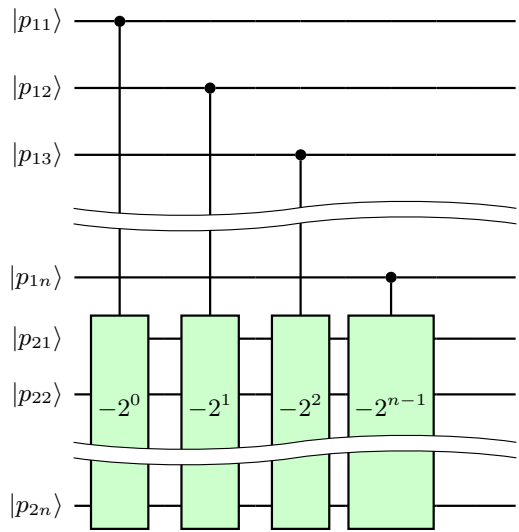}}
    \caption{Quantum circuit scheme for the permutation of terms step in QLBM using Carleman. The green boxes refer to the propagation step, where we have written how much we subtract from the original value of the register in each case. Each propagation unitary is applied for states with $\ket{1}_\tau$}
    \label{fig:scheme_permutation}
\end{figure}

    \item \textbf{Collision:} In general, the collision for LBM is not unitary, which is also the case when we apply Carleman to the embedded quantum state. To make it unitary, we create a collision operator $C$ embedded in the target number of qubits (For D2Q9, four qubits for ${c_1}$, four qubits for $c_2$, one qubit for $\tau$ and one ancilla qubit to encode the diagonal of $g$) and apply a linear combination of unitaries (LCU) \cite{childs2012hamiltonian} in combination with singular value decomposition (SVD) \cite{zhang2015svd}. As a result of SVD, we obtain $C=UDV$ with $D$ a diagonal and non-unitary matrix. Then, we divide $D$ into a sum of unitaries as $D=D_1+D_2$ with $D_1=D+i\sqrt{I-D^2}$ and $D_2=D-i\sqrt{I-D^2}$ to which we can apply LCU. The encoding of a dense unitary operator in the collision register scales as $O(Q^3)$ with $Q$ the number of channels used. On the other hand, at each time step, we need to encode which lattice sites are diagonal $(\ket{x_2}=\ket{0}$). The step requires a multi-controlled Toffoli gate, and without assuming access to additional ancillas, the scaling is $O(log^2(N))$.
    \item \textbf{Propagation :} The propagation step has a computational complexity of $O(log^2(N))$ with the number of lattice sites to propagate. In this case, the propagation is controlled by the collision registers $c_1$ and $c_2$ for $\ket{1}_\tau$ and by ${c_1}$ for $\ket{0}_\tau$. The reason why $\ket{1}_\tau$ is conditioned to both collision registers is due to the specific encoding used.
        \item \textbf{Measurement :} The measurement is the final step of the algorithm. There are many procedures to obtain full tomography of the quantum state \cite{Liu2020VQCTomography,Hwang2023AdaptiveTomography}. However, only a small section of the lattice is of interest in CFD or other physics simulations. In many cases, obtaining the expectation value of a quantity of interest is sufficient. This step will scale as $O(NQ)$ for a general case.

\end{enumerate}

\section{Results and validation}
In this section, we will show results obtained with the quantum circuit presented in section~\ref{sec:quantum} using a quantum emulator in Qiskit \cite{qiskit2024} and validate them classically. The classical validation involves a matrix multiplication using the same embedding. First, we will verify the algorithm for a simple case with $L_x=2$ and five time steps. After that, we will use  $5\cdot 10^8$ shots to evaluate the results accuracy. Notice that the high number of shots is due to the low probability of measuring each time step $10^{-2}$ and the square of the amplitudes that further decrease the relative amplitude of $g$ with respect to $f$.  We will focus on a system with 2 and 4 lattices in each dimension, corresponding to $L_x=4$ and $L_x=16$ lattice sites for $f$ and 16 and 256 for $g$ due to computational limitations. Given the low number of lattice sites, no meaningful CFD simulation can be done. Therefore, to show the capabilities of the quantum algorithm and its accuracy, we will initialise the system randomly (random initialisation of $f$) for $L_x=2$ and $L_x=4$. After that, we will simulate $L_x=32$ and 10 time steps in a classical computer using matrix multiplication but with the same embedded operator as in the quantum algorithm. The test case will be a Taylor-Green vortex, a typical test for CFD. Finally, we will include a comparison between the results with $L_x=8$ and 6 time steps between the quantum algorithm with a statevector simulation and the classical algorithm. For a more complete analysis of the capabilities of Carleman applied to LBM, we refer the reader to \cite{fluids7010024}.

%Maybe we could do this too with the velocity but I do not think is necessary nor adds much value.

\begin{figure}[htbp]
    \centering
    \begin{subfigure}[t]{0.45\textwidth}
        \centering
        \includegraphics[width=\linewidth]{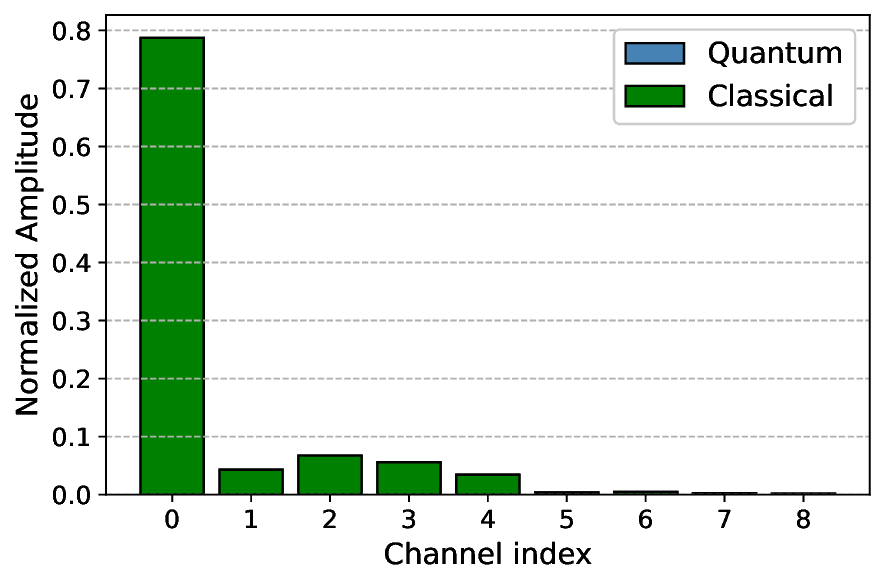}
        \caption{Distribution of channels for $f$ with quantum circuit exact simulation, compared with classical algorithm}
    \end{subfigure}

    \vspace{0.5cm}

    \begin{subfigure}[t]{0.45\textwidth}
        \centering
        \includegraphics[width=\linewidth]{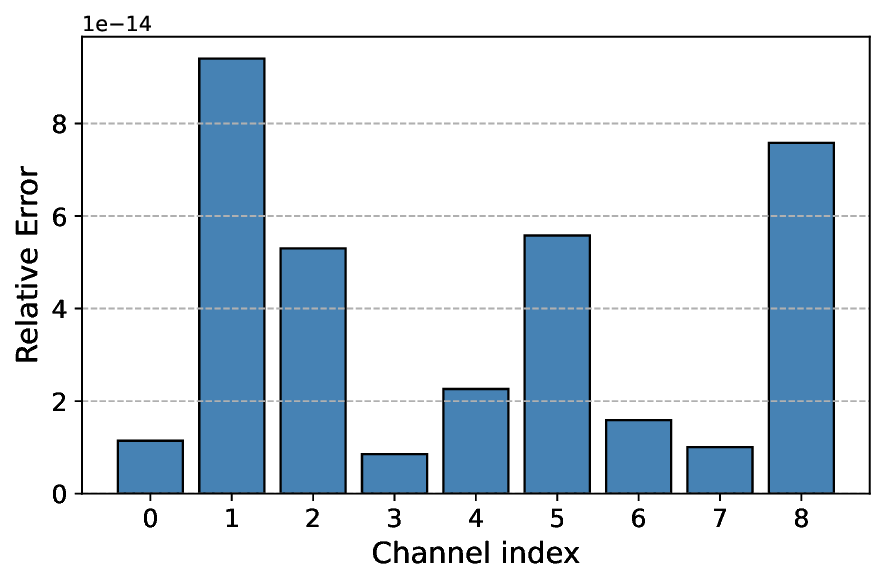}
        \caption{Relative error between classical and quantum results using an exact calculation without shots for $f$}
    \end{subfigure}

    \caption{Distribution of channels for $f$ at $x_1=0$ after 5 time steps using $L_x=2$ and their relative error $\epsilon$ compared to the classical simulation. The notation used here is the standard for LBM D2Q9 following the scheme in Fig~\ref{fig:d2q9}.}
    \label{fig:f_statevector}
\end{figure}

To verify the algorithm's accuracy, we first simulate five time steps with $L_x=2$. In Fig~\ref{fig:f_statevector} we see that the relative error of the statevector simulation with respect to the classical results is minimal, validating the accuracy of the encoding. This small error comes from the limited number of shots used, which adds uncertainty.

\begin{figure}[htbp]
    \centering
    \begin{subfigure}[t]{0.45\textwidth}
        \centering
        \includegraphics[width=\linewidth]{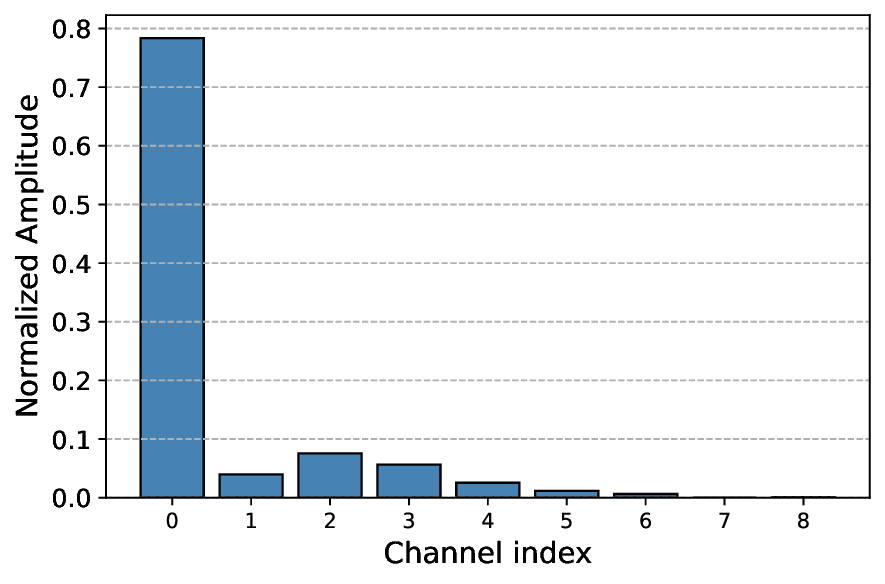}
        \caption{Distribution of channels for $f$ with quantum circuit}
    \end{subfigure}

    \vspace{0.5cm}

    \begin{subfigure}[t]{0.45\textwidth}
        \centering
        \includegraphics[width=\linewidth]{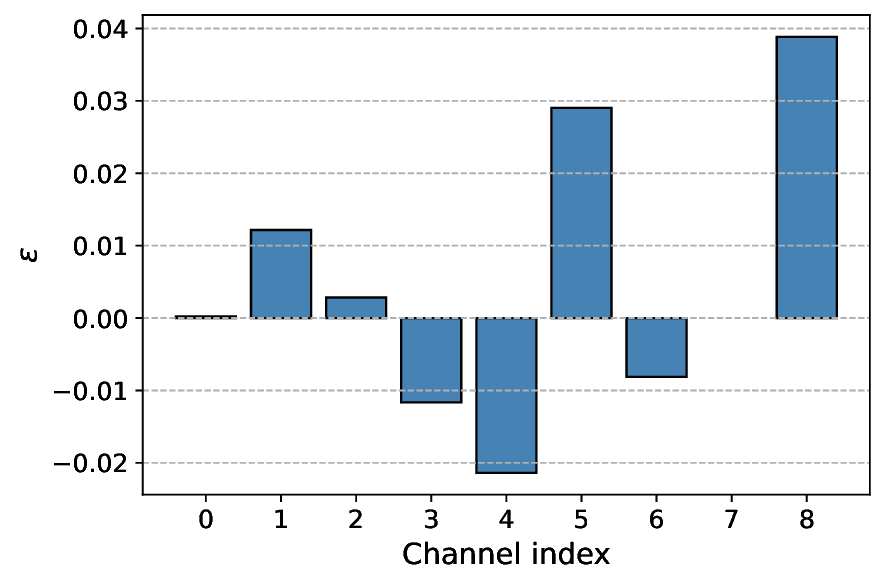}
        \caption{Relative error between classical and quantum circuit for $f$}
    \end{subfigure}

    \caption{Distribution of channels for $f$ at $x_1=0$ after 2 time steps using $L_x=2$ and their relative error compared to the classical simulation. The notation used here is the standard for LBM D2Q9 following the scheme in Fig~\ref{fig:d2q9}. }
    \label{fig:f_one_site}
\end{figure}

\begin{figure}[htbp]
    \centering
    \begin{subfigure}[t]{0.45\textwidth}
        \centering
        \includegraphics[width=\linewidth]{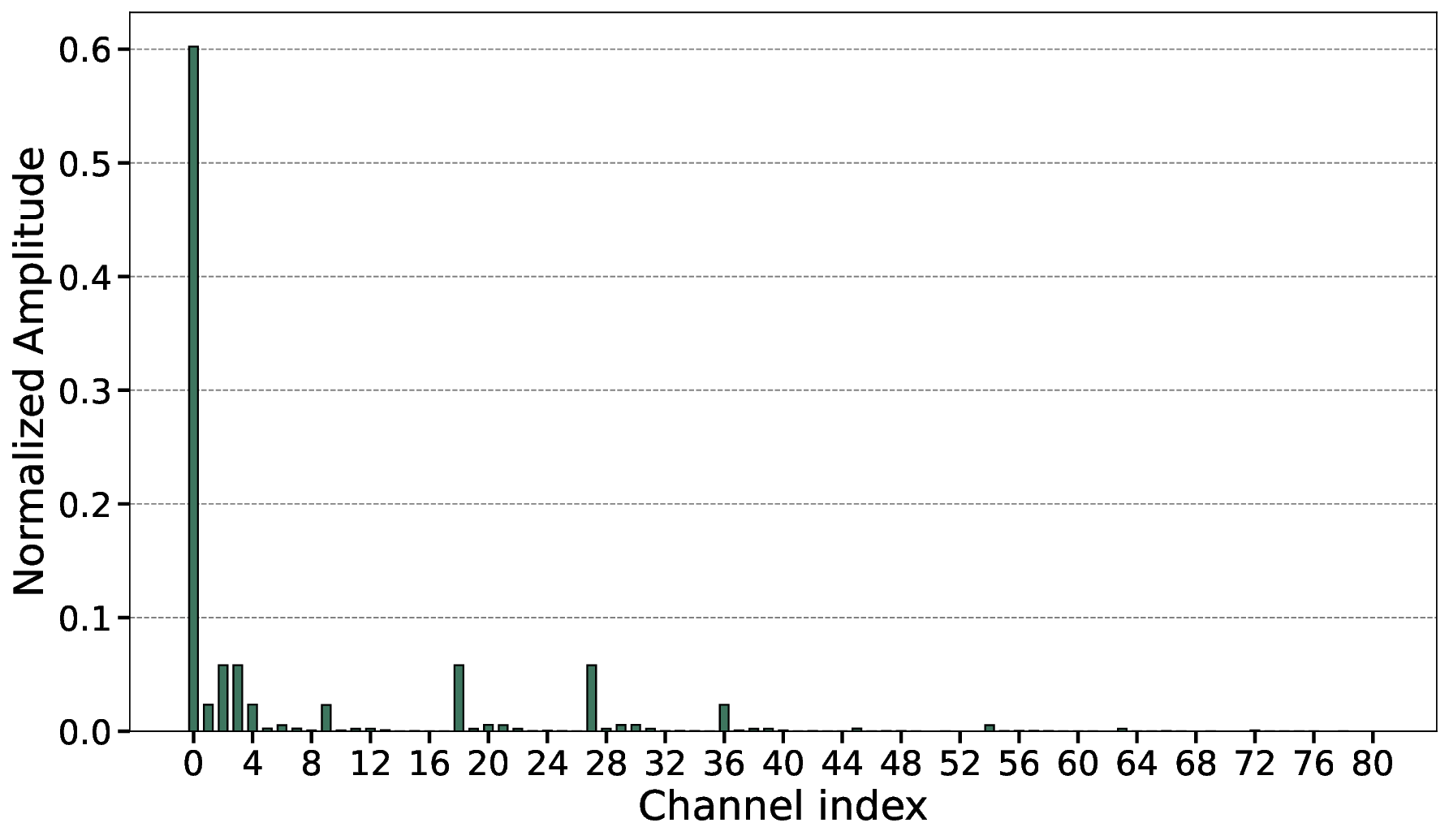}
        \caption{Distribution of channels for $g$ with quantum circuit}
    \end{subfigure}

    \vspace{0.5cm}

    \begin{subfigure}[t]{0.45\textwidth}
        \centering
        \includegraphics[width=\linewidth]{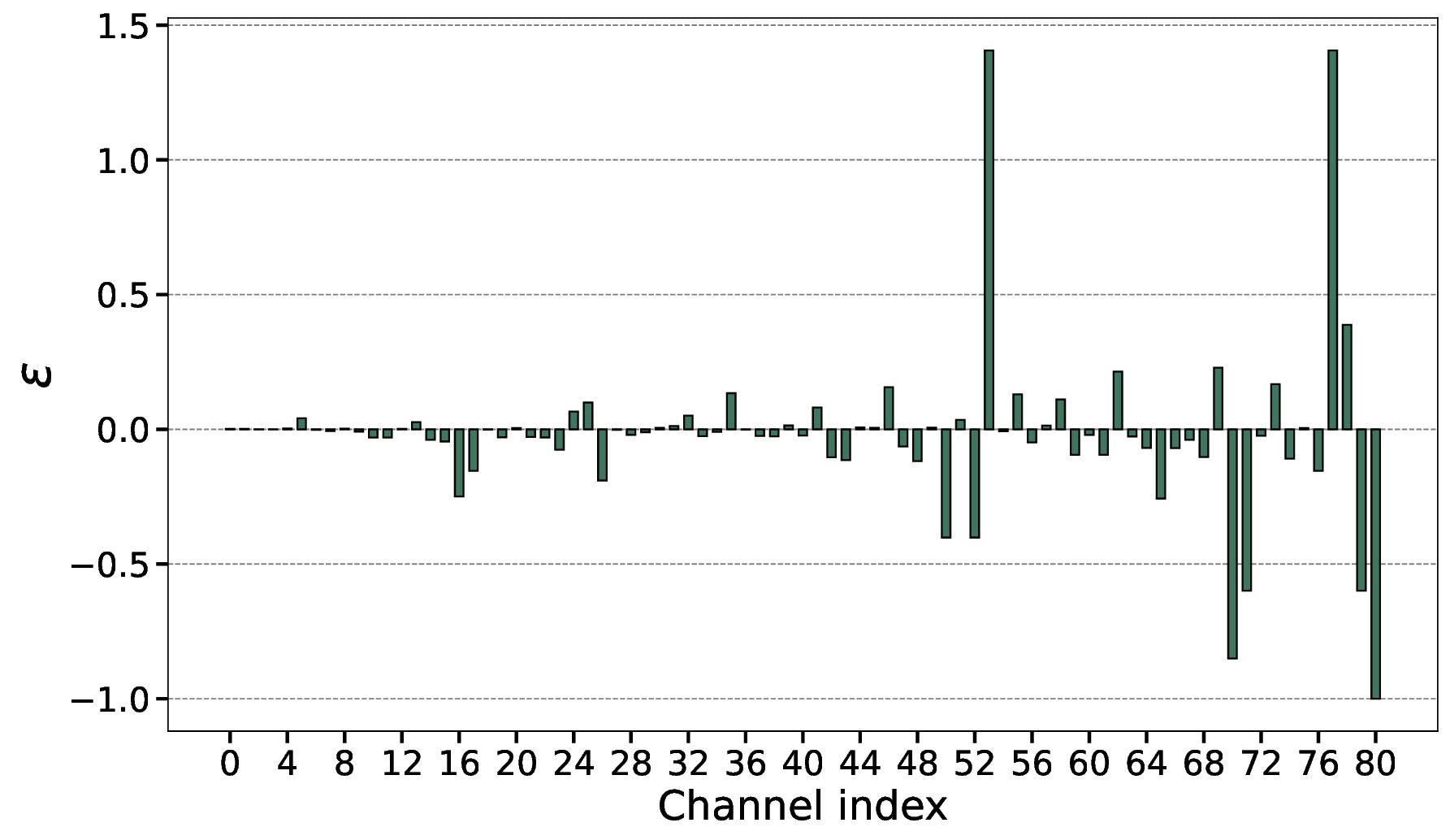}
        \caption{Relative error between classical and quantum circuit for $g$}
    \end{subfigure}

    \caption{Distribution of channels for $g$ at $x_1,x_2=0$ after 2 time steps using $L_x=2$ and their relative error compared to the classical simulation. The channel index refers to the elements $c=(c_1,c_2)$ with the standard for LBM D2Q9 following the scheme in Fig~\ref{fig:d2q9}, where the channels have been arranged in an array with row ordering. }
    \label{fig:g_one_site}
\end{figure}

Now, we will compare how the algorithm behaves when using shots instead of the exact quantum circuit emulator. In Fig~\ref{fig:f_one_site} and Fig~\ref{fig:g_one_site}, we can see the results for $f$ and $g$ and the relative error compared with the classical results. While for $f$ we see small errors (lower than 5\%), for $g$ we observe higher relative errors. The reason for this is that $g$ has a relative probability 100 times smaller than $f$. Specifically, the channels 5-8 in $f$ have a very low probability, which makes specific channels for $g$ very small and therefore a minor error has a significant impact in $\epsilon$ for the corresponding channels.

\begin{figure}[htbp]
    \centering
    \begin{subfigure}[t]{0.45\textwidth}
        \centering
        \includegraphics[width=\linewidth]{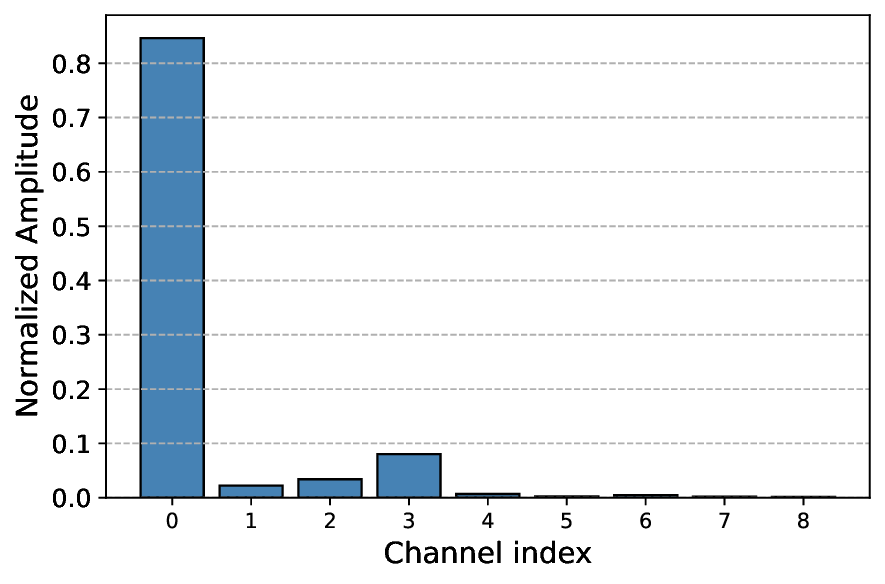}
        \caption{Distribution of channels for $f$ with quantum circuit}
    \end{subfigure}

    \vspace{0.5cm}

    \begin{subfigure}[t]{0.45\textwidth}
        \centering
        \includegraphics[width=\linewidth]{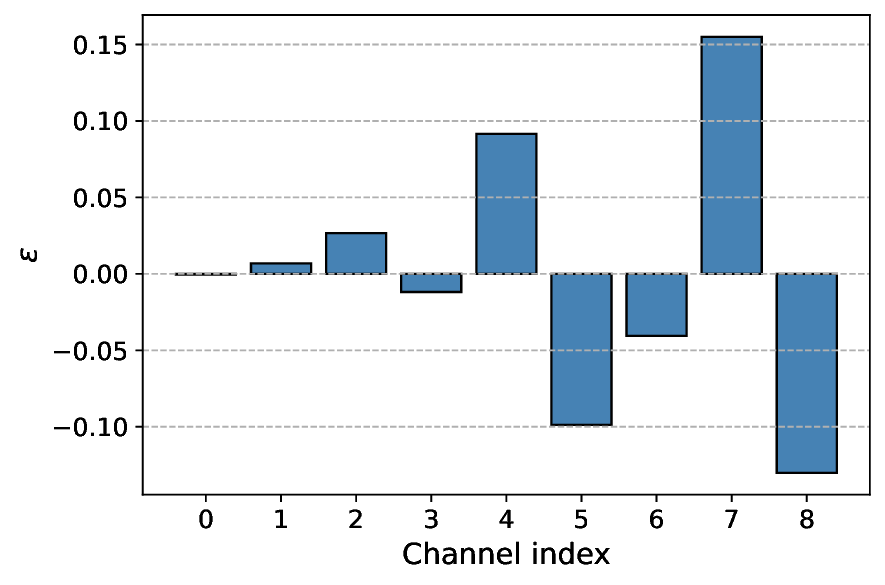}
        \caption{Relative error between classical and quantum circuit for $f$}
    \end{subfigure}

    \caption{Distribution of channels for $f$ at $x_1=0$ after 2 time steps using $L_x=4$ and their relative error $\epsilon$ compared to the classical simulation. The notation used here is the standard for LBM D2Q9 following the scheme in Fig~\ref{fig:d2q9}.}
    \label{fig:f_one_site_4}
\end{figure}

\begin{figure}[htbp]
    \centering
    \begin{subfigure}[t]{0.45\textwidth}
        \centering
        \includegraphics[width=\linewidth]{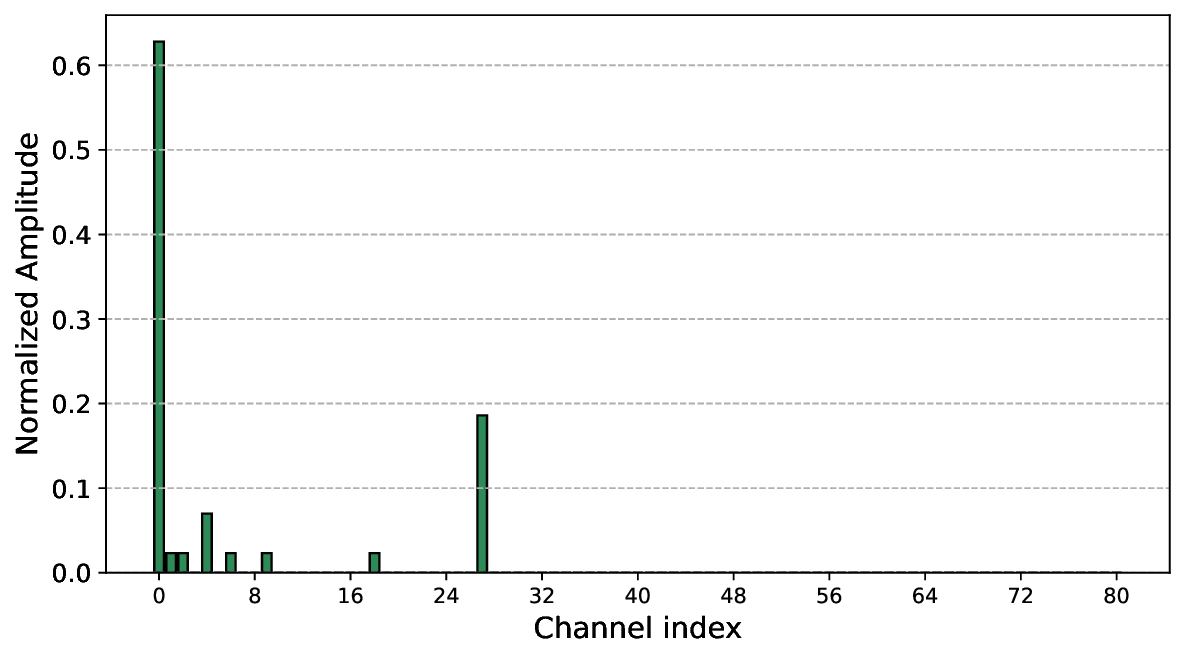}
        \caption{Distribution of channels for $g$ with quantum circuit}
    \end{subfigure}

    \vspace{0.5cm}

    \begin{subfigure}[t]{0.45\textwidth}
        \centering
        \includegraphics[width=\linewidth]{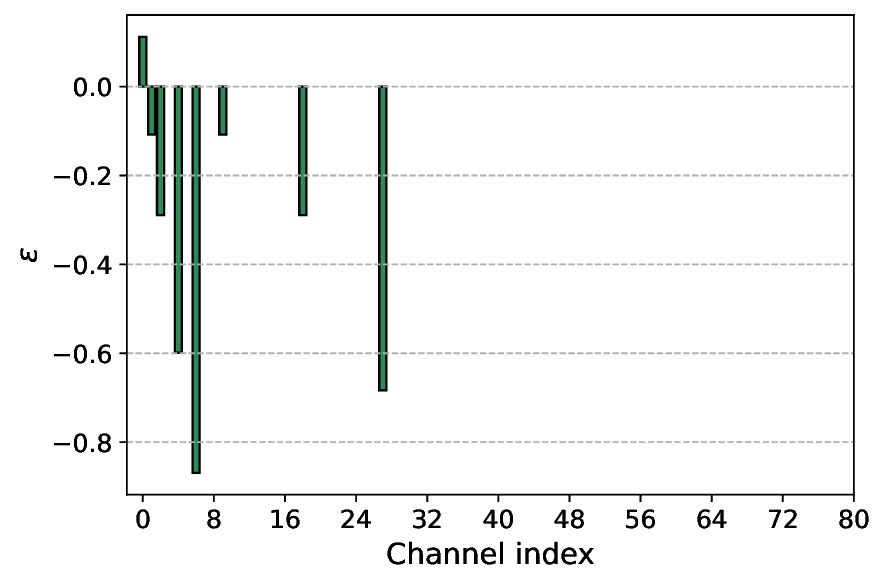}
        \caption{Relative error between classical and quantum circuit for $g$}
    \end{subfigure}

    \caption{Distribution of channels for $g$ at $x_1,x_2=0$ after 2 time steps using $L_x=4$ and their relative error compared to the classical simulation. The channel index refers to the elements $c=(c_1,c_2)$ with the standard for LBM D2Q9 following the scheme in Fig~\ref{fig:d2q9}, where the channels have been arranged in an array with row ordering. }
    \label{fig:g_one_site_4}
\end{figure}

If we do the same comparison again for a larger lattice site $L_x=4$, we will obtain higher relative errors as seen in Fig~\ref{fig:f_one_site_4}. This increment in $\epsilon$ is due to the higher number of variables encoded, which decreases the amplitudes at the state. As Fig~\ref{fig:g_one_site_4} reveals, for $g$ the relative errors are also very high in channels with a high probability of measuring (within that lattice), which means that the number of shots is insufficient.

\begin{figure}[htbp]
    \centering
    \begin{subfigure}[t]{0.45\textwidth}
        \centering
        \includegraphics[width=\linewidth]{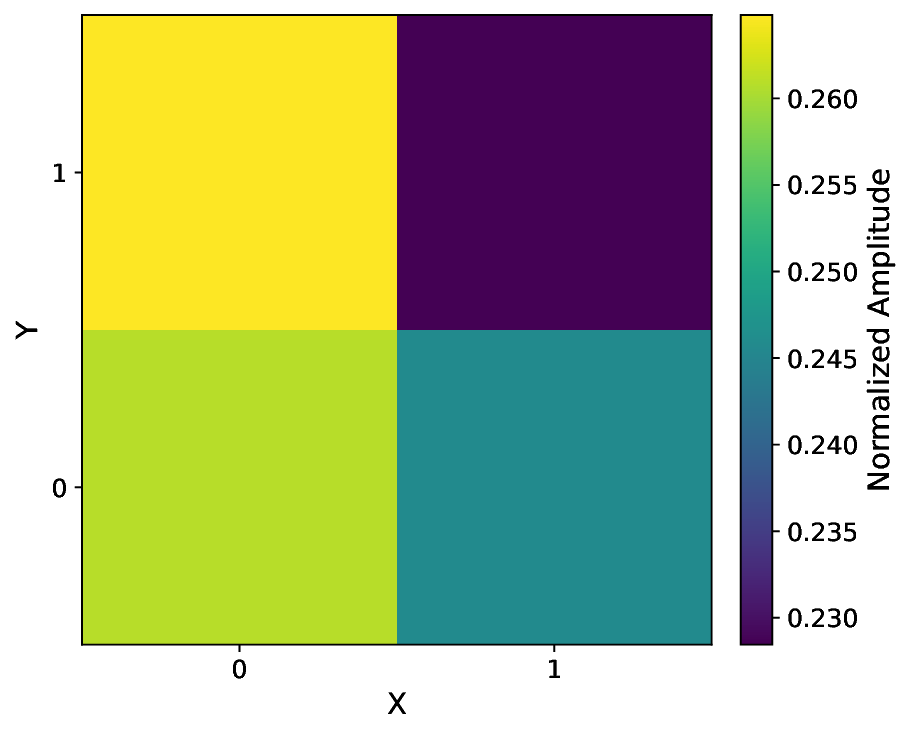}
        \caption{Amplitude for each lattice site for $\sigma_f$ with quantum circuit simulation}
    \end{subfigure}

    \vspace{0.5cm}

    \begin{subfigure}[t]{0.45\textwidth}
        \centering
        \includegraphics[width=\linewidth]{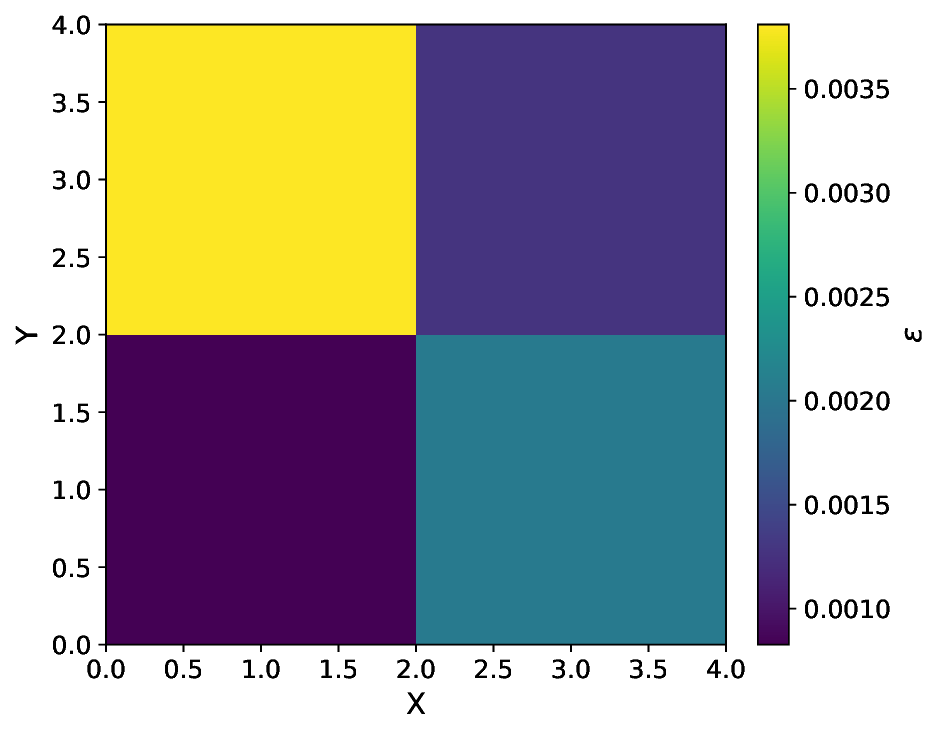}
        \caption{Relative error between classical and quantum circuit for $\sigma_f$ at each lattice site}
    \end{subfigure}

    \caption{Distribution of total amplitude for different lattice sites for $\sigma_f$ after 2 time steps using a 2D lattice with $L_x=2$ and their relative error $\epsilon$ compared to the classical simulation.} 
    \label{fig:f_all_site_2}
\end{figure}

\begin{figure}[htbp]
    \centering
    \begin{subfigure}[t]{0.45\textwidth}
        \centering
        \includegraphics[width=\linewidth]{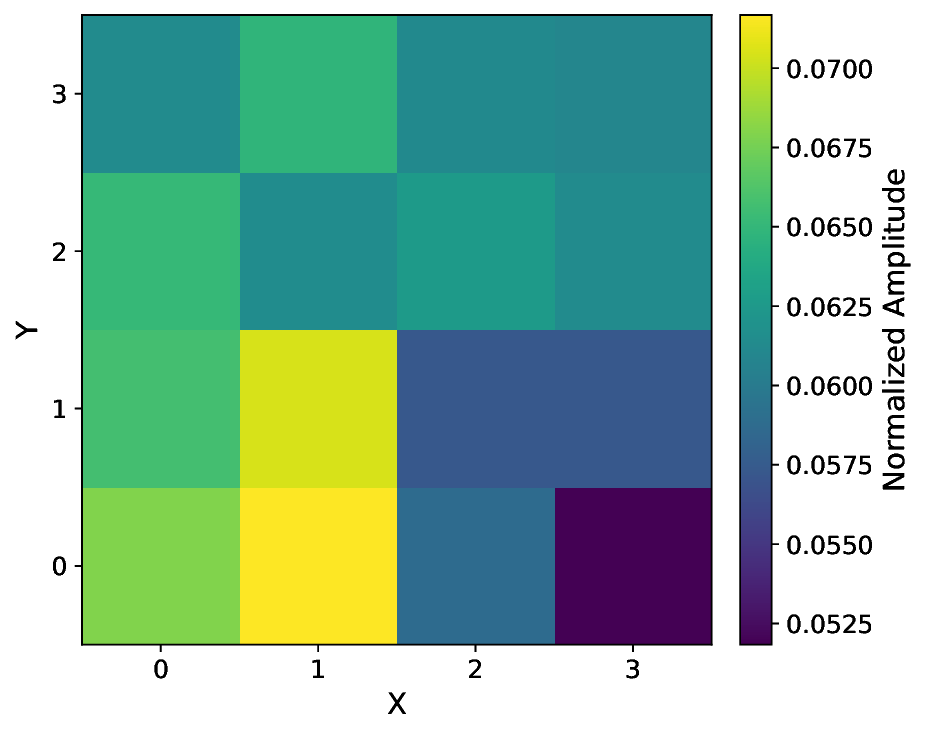}
        \caption{Amplitude for each lattice site for $\sigma_g$ with quantum circuit simulation}
    \end{subfigure}

    \vspace{0.5cm}

    \begin{subfigure}[t]{0.45\textwidth}
        \centering
        \includegraphics[width=\linewidth]{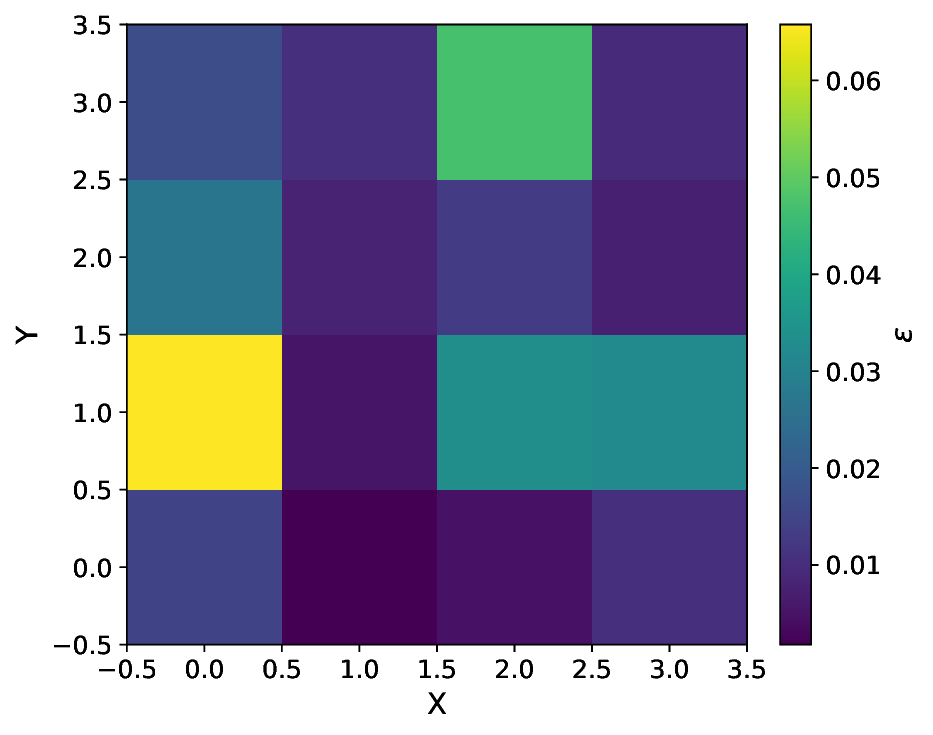}
        \caption{Relative error between classical and quantum circuit for $\sigma_g$ at each lattice site}
    \end{subfigure}

    \caption{Distribution of total amplitude for different lattice sites in $\sigma_g$ after 2 time steps using $L_x=2$ and their relative error $\epsilon$ compared to the classical simulation.}
    \label{fig:g_all_site_2}
\end{figure}

Now, we will do the same comparison using quantum circuit emulation with shots and classical simulations for macroscopic variables (density). We will measure the mass of each lattice site for $L_x=2$ after two time steps. As the total probability for each lattice site is much higher than the individual probability for each channel, we should observe better results. For the notation we will use $\sigma_f(x)=\sum_i f_i(x)$ and $\sigma_g(x_1,x_2)=\sum_{i,j} g_{ij}(x_1,x_2)$. In Fig~\ref{fig:f_all_site_2} and Fig~\ref{fig:g_all_site_2}, we can observe how the amplitudes match very well, with minimal errors after 2 time steps. 

%If we want to save space and put less figures, we could delete the amplitude for f and g here nad only let the one for relative errors 

\begin{figure}[htbp]
    \centering
    \begin{subfigure}[t]{0.45\textwidth}
        \centering
        \includegraphics[width=\linewidth]{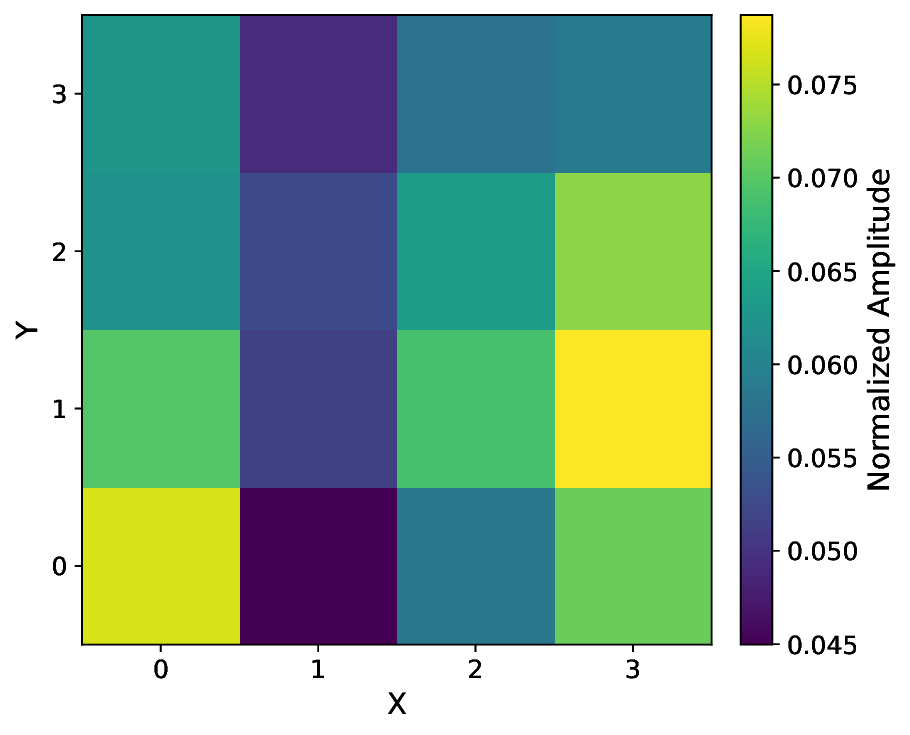}
        \caption{Amplitude for each lattice site in $\sigma_f$ with quantum circuit simulation}
    \end{subfigure}

    \vspace{0.5cm}

    \begin{subfigure}[t]{0.45\textwidth}
        \centering
        \includegraphics[width=\linewidth]{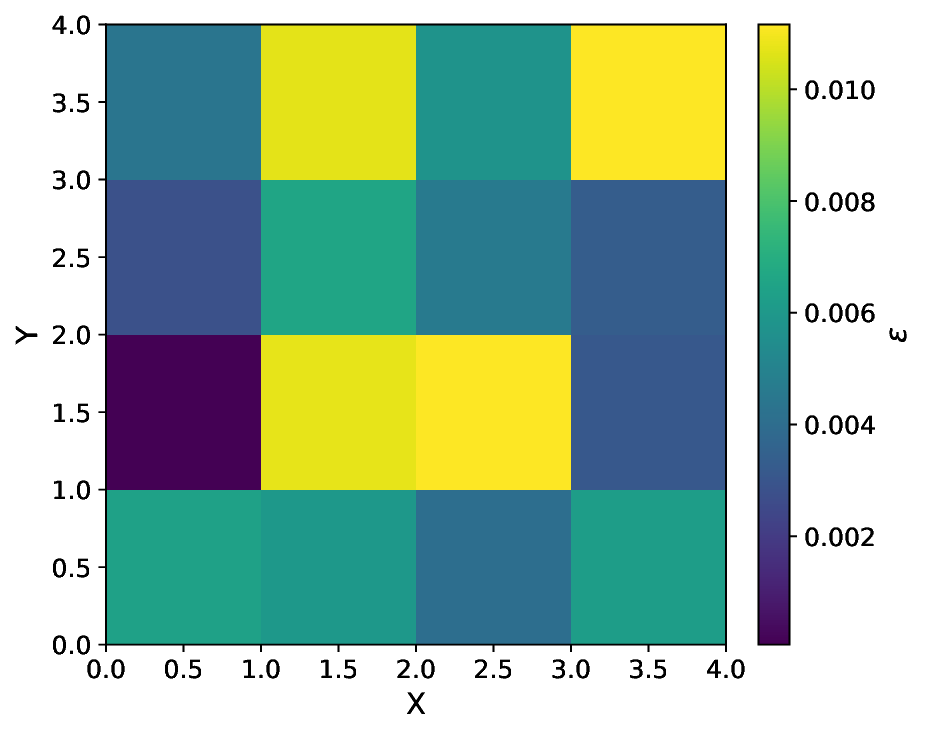}
        \caption{Relative error between classical and quantum circuit for $\sigma_f$ at each lattice site}
    \end{subfigure}

    \caption{Distribution of total amplitude for different lattice sites in $\sigma_f$ after 2 time steps using $L_x=4$ and their relative error $\epsilon$ compared to the classical simulation.}
    \label{fig:f_all_site_4}
\end{figure}

\begin{figure}[htbp]
    \centering
    \begin{subfigure}[t]{0.45\textwidth}
        \centering
        \includegraphics[width=\linewidth]{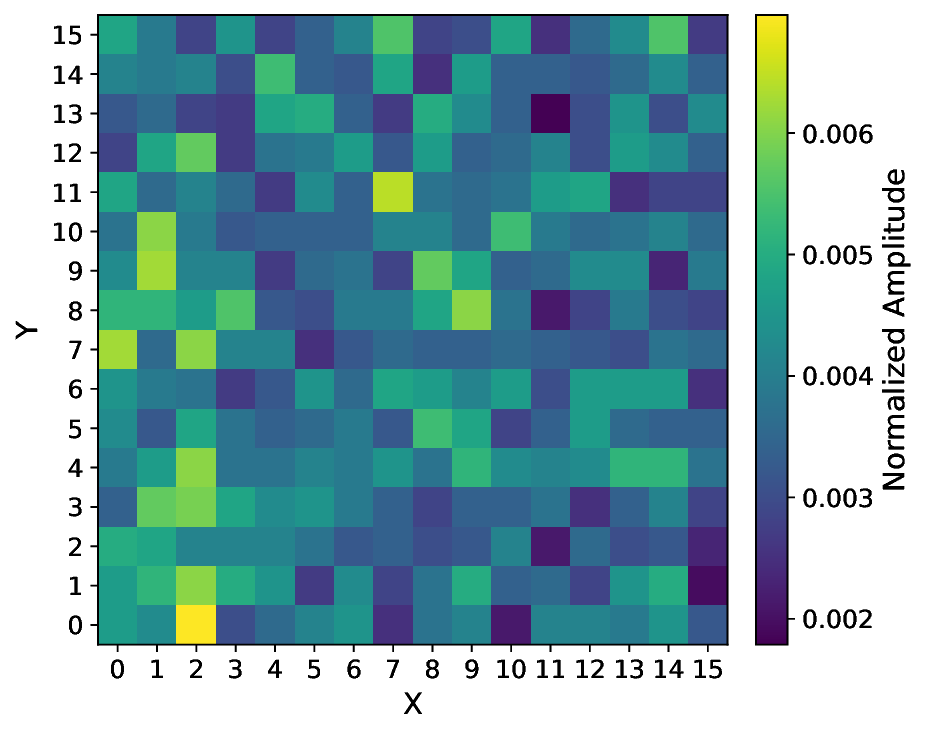}
        \caption{Amplitude for each lattice site in $\sigma_g$ with quantum circuit simulation}
    \end{subfigure}

    \vspace{0.5cm}

    \begin{subfigure}[t]{0.45\textwidth}
        \centering
        \includegraphics[width=\linewidth]{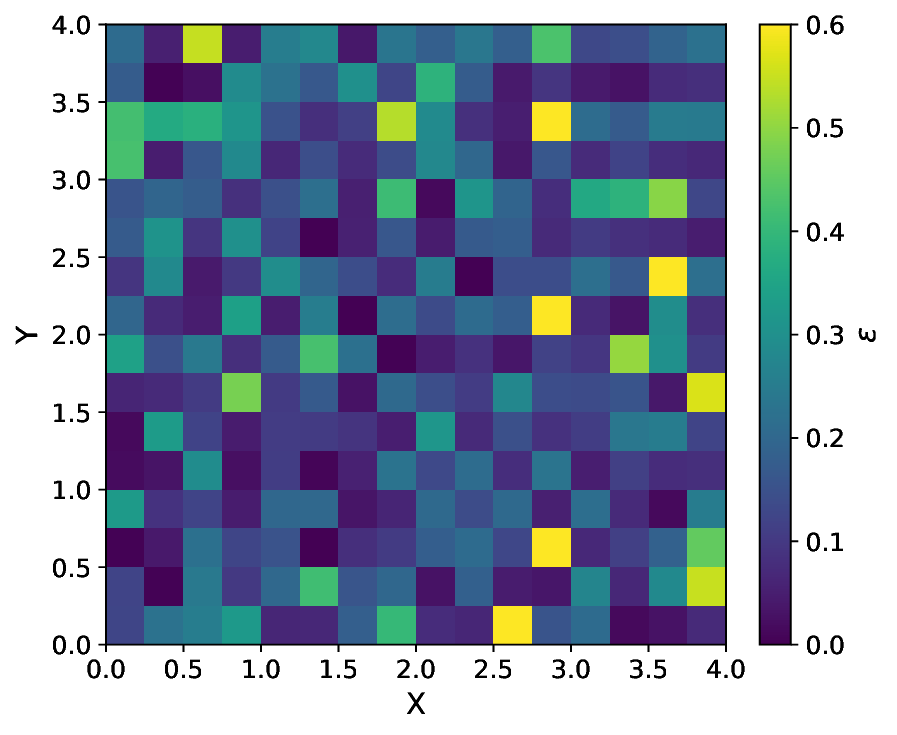}
        \caption{Relative error between classical and quantum circuit for $\sigma_g$ at each lattice site}
    \end{subfigure}

    \caption{Distribution of total amplitude for different lattice sites in $\sigma_g$ after 2 time steps using $L_x=4$ and their relative error $\epsilon$ compared to the classical simulation. The figure compares Carleman linearization with LBM using a classical simulation.}
    \label{fig:g_all_site_4}
\end{figure}

As was the case with the distribution of channels, when we increase the number of lattice sites for $L_x=4$, we obtain higher relative errors due to having lower amplitudes at each state. This behaviour can be seen in Fig~\ref{fig:f_all_site_4} and  Fig~\ref{fig:g_all_site_4}. As previously seen for the distribution of channels, the probability of measuring each channel of $g$ is very low, and the relative error is also high when we sum the terms at each lattice site.

\begin{figure}[htbp]
    \centering
    \begin{subfigure}[t]{0.45\textwidth}
        \centering
        \includegraphics[width=\linewidth]{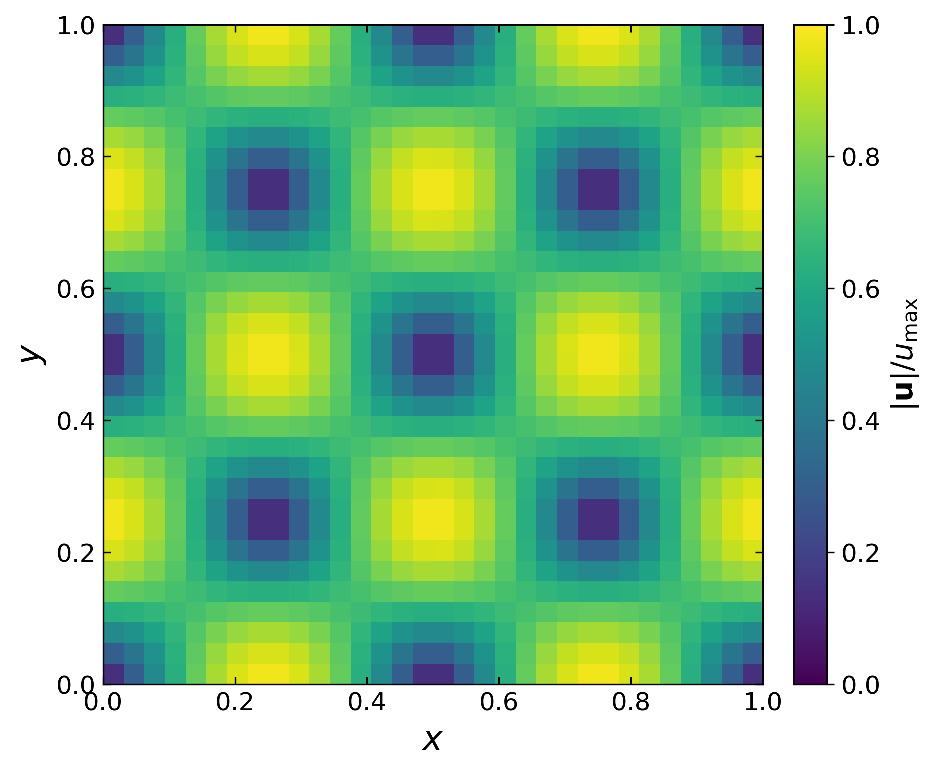}
        \caption{Initial velocity field}
    \end{subfigure}

    \vspace{0.5cm}

    \begin{subfigure}[t]{0.45\textwidth}
        \centering
        \includegraphics[width=\linewidth]{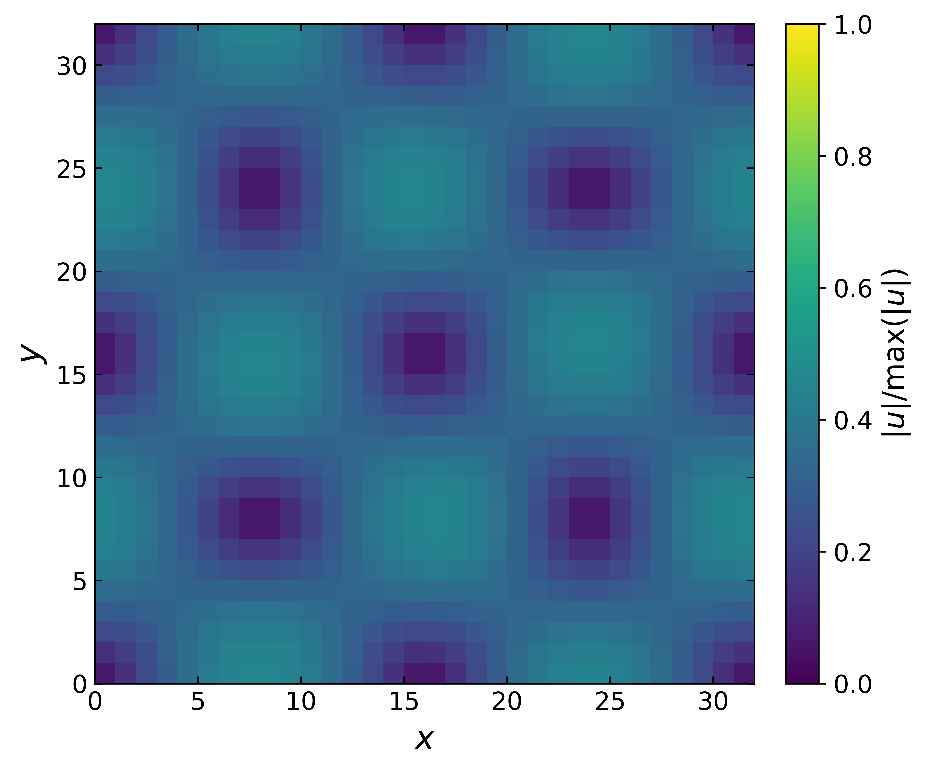}
        \caption{Velocity field using LBM with Carleman linearization after 10 time steps}
    \end{subfigure}

    \begin{subfigure}[t]{0.45\textwidth}
        \centering
        \includegraphics[width=\linewidth]{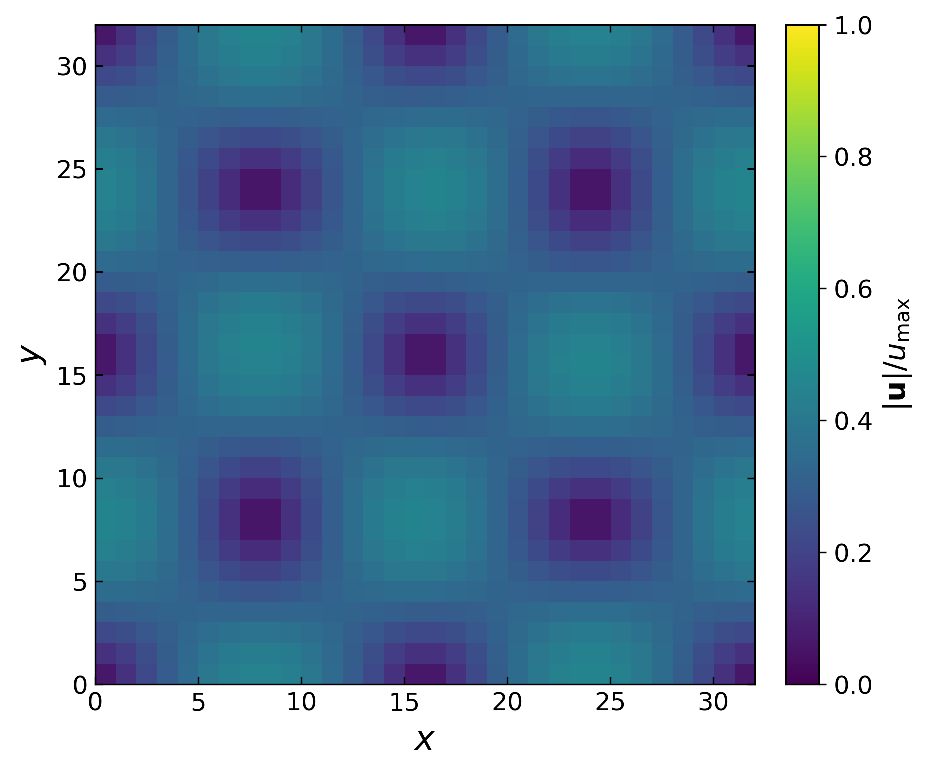}
        \caption{Velocity field using LBM after 10 time steps}
    \end{subfigure}

    \caption{Velocity field of a Taylor-Green vortex simulation using LBM with Carleman linearization after 10 time steps compared to LBM with $u_{max}=0.15$ and $\tau=5$.}
    \label{fig:lbm_tgv}
\end{figure}

\begin{figure}[htbp]
    \centering
    \begin{subfigure}[t]{0.45\textwidth}
        \centering
        \includegraphics[width=\linewidth]{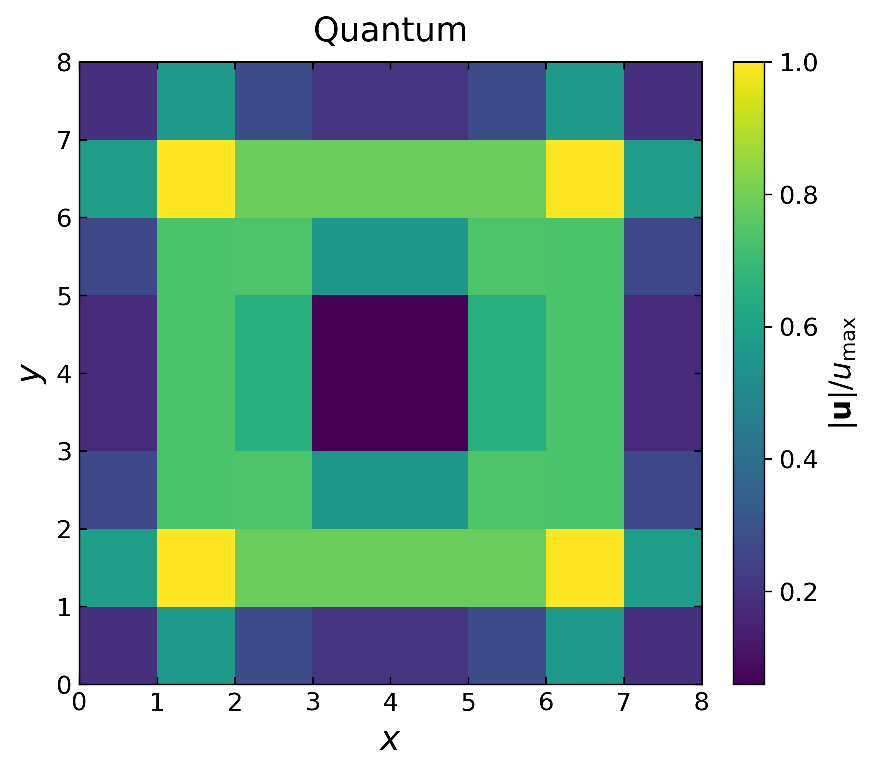}
    \end{subfigure}

    \vspace{0.5cm}

    \begin{subfigure}[t]{0.45\textwidth}
        \centering
        \includegraphics[width=\linewidth]{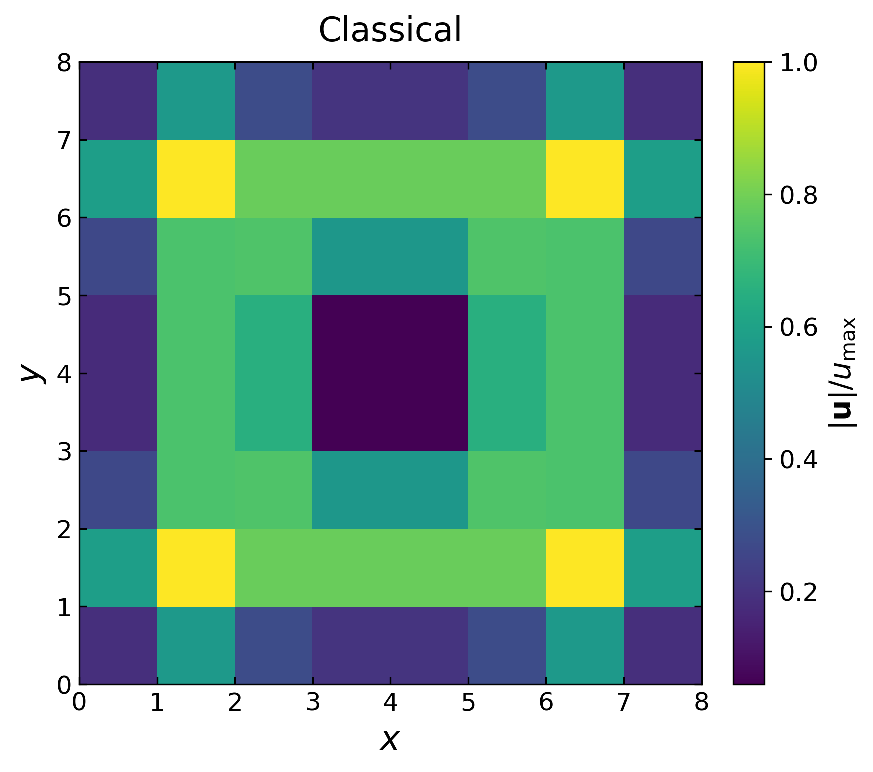}
    \end{subfigure}

    \begin{subfigure}[t]{0.45\textwidth}
        \centering
        \includegraphics[width=\linewidth]{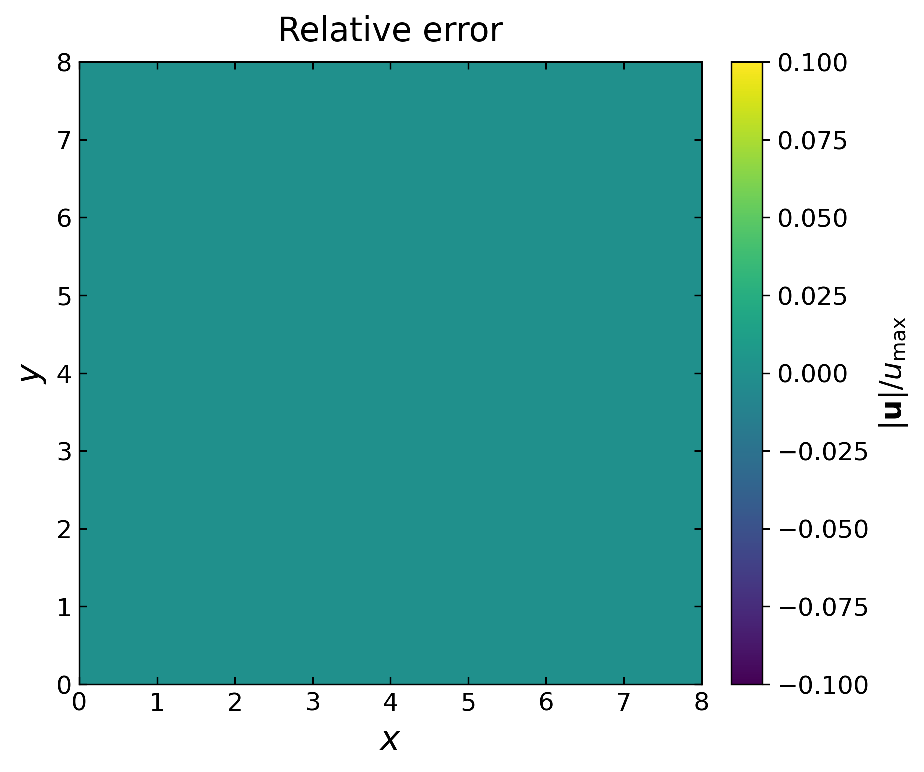}
    \end{subfigure}

    \caption{Velocity field of a Taylor-Green vortex simulation using LBM with Carleman linearization after 6 time steps with $u_{max}=0.15$ and $\tau=5$. The figure compares a quantum circuit simulation with statevector and a classical simulation.}
    \label{fig:lbm_tgv_q}
\end{figure}

From the analysed results, we can conclude that the model works well, obtaining accurate results using a quantum circuit for LBM with Carleman for two time steps and more. However, the low probability to measure each time step due to the usage of LCU produces high errors in the extraction of the results when using $5\cdot 10^8$ shots (which is the maximum number of shots before running out of memory). To end this section and complement the analysis, we will make use of the collision operator encoded in the quantum algorithm to simulate a simple CFD case for a higher number of lattice sites, in this case $L_x=32$ after 10 time steps (Fig~\ref{fig:lbm_tgv}). The figure shows the same velocity field after 10 time steps between the LBM simulation and the linearized with Carleman (classically) using the same collision operator. Notice that the method used for linearization takes into account that $\rho\approx 1$ (weakly compressible limit), which may explain subtle differences in the results.
Additionally, the second order is computed only using the operator $A\otimes A$, which is a linear approximation. To compare the difference between the quantum circuit results with a statevector simulation and the classical algorithm for a use case, we used $L_x=8$ and six time steps due to computational limitations. The initial state is the same as for the case shown in Fig~\ref{fig:lbm_tgv}, with a shorter characteristic wave length of $k_x=\pi / Lx$. In Fig~\ref{fig:lbm_tgv_q}, we can see that the error after six time steps is zero, showing that we are doing the exact simulation in the quantum circuit as in the classical simulation.

\section{Conclusions}
In this paper we have developed a novel encoding for quantum Lattice Boltzmann Method (QLBM) using Carleman linearization, focusing on second order. In contrast to other approaches such as \cite{Sanavio2024LatticeBC}, where the collision is not local and the depth scales as $O(N^4Q^4)$ and \cite{sanavio2025carleman}, where the collision does not scale with the lattice sites but the success probability at each time step is very low, of the order of $10^{-6}$, our algorithm has higher probability and locality. 
Our method allows to obtain a QLBM algorithm for several time steps by making the method local. Each algorithm step was carefully reviewed and the computational complexity derived, obtaining a scaling of $O(log_2^3(N)+T(log_2^2(N)+Q^3))$ with $N$ the number of lattice sites and $Q$ the number of channels and $T$ the number of time steps. Despite the advancements made in this paper, more challenges are yet to be addressed to make the algorithm competitive and obtain a real quantum advantage.
First, the probability to measure each time-step is dependent on the initial state and the relaxation time from LBM equations due to LCU linearization. According to our tests for a lattice with $N=16$, the probability of measuring the right result is of the order of $10^{-2}$. Second, despite being local, the collision operator is still very demanding to encode with a considerable depth using Qiskit's built-in unitary operator decomposition \cite{qiskit2024}. Both of these problems are associated with the non-unitarity of the operator and alternative approximations or methods to make the operator unitary may be needed.

\FloatBarrier

\appendix
\section{Permutation of states and quantum algorithm proof}
\label{sec:apx}

In this section, we will prove the validity of the encoding and analyze how the quantum state is modified after each operator in the algorithm. First, we will prove that the encoding is the correct one. For that we need to show that using the following encoding 

\begin{equation}
\begin{aligned}
    c_v(x_1, x_2) &= c_v(x_{11}, x_{12}, x_{21}, x_{22}) \\
                  &= g\left(x_{11}, x_{12},\; [x_{11} + x_{21}] \bmod L_x,\; [x_{12} + x_{22}] \bmod L_y\right) \\
                  &= g(x_1, \tilde{x}_2)
\end{aligned}
\end{equation}

where $c_v$ is an auxiliary variable for the encoding, $g$ is the Carleman second-order dynamical variable, $x_1=(x_{11},x_{12})$ and $x_2=(x_{21},x_{22})$ . When $x_1=\tilde{x}_2$, then $g(x_1,x_1)$ is stored in  $c_v(x_1,0)$, obtaining

\begin{equation}
\begin{aligned}
    x_{11}&=[x_{11}+x_{21}]\bmod{L_x}\rightarrow x_{21}=0\\
    x_{12}&=[x_{12}+x_{22}]\bmod{L_y}\rightarrow x_{22}=0
\end{aligned}
\end{equation}

Therefore, we proved that when  $x_1=\tilde{x}_{2}$, our particular encoding ensures that the value is stored in $c_v(x_1,0)$.
The next step is to show that this manipulation does not require a non-local propagation for the channels caused by the rearrangement of terms in $c_v$. When shifting the second variable $x_2$ from $g_{ij}(x_1,x_2)$, we get that for the variable with the new encoding $c_v$

\begin{equation}
\begin{aligned}
c_v(x_1, x_2 + \tilde{u}_{c_2}) =\ 
& g\Big( x_{11}, x_{12},\ 
      [x_{11} + x_{21} + \tilde{u}_{c_{21}}] \bmod L_x,\ \\
&\quad [x_{12} + x_{22} + \tilde{u}_{c_{22}}] \bmod L_y \Big)
\end{aligned}
\end{equation}

where we have used the notation $\tilde{u}$ for the direction of propagation depending on the channel ($\pm 1)$, with the first index one for $x_1$ and two for $x_2$ and the second subindex to the component (two components as we are working in 2D). For the initial variable $g$, we have that

\begin{equation}
g(x_1,x_2+u_{c_2})=g\left(x_{11}, x_{12}, [x_{21}+u_{c_{21}}] \bmod L_x, [x_{22}+u_{c_{22}}] \bmod L_x\right)
\end{equation}

to obtain the same increment in both of them, $\tilde{u_2}=u_2=(0,0,u_{21},u_{22})$

Now due to the first dimension being couple with the second dimension, we have that

\begin{equation} 
\begin{aligned}
c_v(x_1 + \tilde{u}_{c_1}, x_2) 
&= g\Big( 
[x_{11} + \tilde{u}_{c_{11}}] \bmod L_x,\ 
[x_{12} + \tilde{u}_{c_{12}}] \bmod L_y,\\
&\quad [x_{11} + x_{21} + \tilde{u}_{c_{11}}] \bmod L_x,\\
&\quad[x_{12} + x_{22} + \tilde{u}_{c_{12}}] \bmod L_y 
\Big)
\end{aligned}
\end{equation}

while for $g$ we have 

\begin{equation} 
\begin{aligned}
g(x_1 + u_{c_1}, x_2) 
&= g\Big( 
[x_{11} + u_{c_{11}}] \bmod L_x,\\ 
&\qquad[x_{12} + u_{c_{12}}] \bmod L_y, x_{21},\ x_{22} 
\Big)
\end{aligned}
\end{equation}

This means that we have $\tilde{u}_{c_1}$=($u_{c_{11}}$,$u_{c_{12}}$,-$u_{c_{11}}$,-$u_{c_{12}}$) to match the shift for the original variable, where the first and second components do not alter the third and fourth. This also proves that the encoding in the algorithm can be done only once at the beginning, as the lattice order is preserved during the computation, encoding the element $g(x_1,y_{aux})$ in the Carleman variable $c_v(x_1,x_2)$ permanently. When the simulation finish, the original encoding can be recovered. 

Now, let's suppose that the combination of $A,B$ and $A\otimes A$ are unitary. First, we will show how each operator transforms the state vector, and then, we will show their application within one operator as done in the algorithm. For the quantum state, we will use the notation $\ket{x_1,x_2,i,j,0/1}$ indicating the first lattice, the second lattice, the first channel, the second channel and the nonlinear order quantum registers. Then, if the initial state after state preparation is

\begin{equation}
\ket{\Psi}=\sum_{p_1,p_2}^N \sum_{i,j}^C \sum_{\tau}^2{ c_v(x_1,x_2,i,j,\tau)\ket{x_1,x_2,i,j,0/1}}
\end{equation}

where we used $c_v$ to refer to the embedding of $f$ and $g$, we have 

\begin{equation}
\begin{aligned}
A\ket{\Psi} =\ 
& \sum_{x_1,x_2=1}^{N} \sum_{i,j=1}^{C} \beta_{x_{1_{i}} x_{2_{j}}} \ket{x_1,x_2,i,j,1} \\
& + \sum_{x_1=1}^{N} \sum_{i,l=1}^{C} A_{il}\, \alpha_{x_i} \ket{x_1,0,i,0,0} \\
=\ 
& \sum_{x_1,x_2=1}^{N} \sum_{i,j=1}^{C} \beta_{x_{1_{i}} x_{2_{j}}} \ket{x_1,x_2,i,j,1} \\
& + \sum_{x_1=1}^{N} \sum_{i=1}^{C} \tilde{\alpha}_{x_i} \ket{x_1,0,i,0,0}
\end{aligned}
\end{equation}

For $B$ we have

\begin{equation}
\begin{aligned}
B\ket{\Psi} \ket{a} =\ 
& \sum_{x_1=1}^{N} \sum_{i,j=1}^{C} B_{ijkl}\, \beta_{x_i 0_j} \ket{x_1,0,i,j,1} \ket{1} \\
& + \sum_{\substack{x_1,x_2=1 \\ x_2 \ne 0}}^{N} \sum_{i,j,k,l=1}^{C} \beta_{x_i y_j} \ket{x_1,x_2,i,j,1} \ket{0} \\
& + \sum_{x_1=1}^{N} \sum_{i=1}^{C} \alpha_{x_i} \ket{x_1,0,i,0,0} \ket{1} \\
=\ 
& \sum_{\substack{x_1,x_2=1 \\ x_2 \ne 0}}^{N} \sum_{i,j=1}^{C} \beta_{x_i y_j} \ket{x_1,x_2,i,j,1} \ket{0} \\
& + \sum_{x_1=1}^{N} \sum_{i=1}^{C} \left( \alpha_{x_i} + \tilde{\alpha}_{x_i} \right) \ket{x_1,0,i,0,0} \ket{1}
\end{aligned}
\end{equation}
And for $A\otimes A$ we obtain
\begin{equation}
\begin{aligned}
A \otimes A \ket{\Psi} =\ 
& \sum_{x_1,x_2=1}^{N} \sum_{i,j,k,l=1}^{C} A_{ijkl}\, \beta_{x_i y_j} \ket{x_1,x_2,i,j,1} \\
& + \sum_{x_1=1}^{N} \sum_{i=1}^{C} \alpha_{x_i} \ket{x_1,0,i,0,0} \\
=\ 
& \sum_{x_1,x_2=1}^{N} \sum_{i,j=1}^{C} \tilde{\beta}_{x_i y_j} \ket{x_1,x_2,i,j,1} \\
& + \sum_{x_1=1}^{N} \sum_{i=1}^{C} \alpha_{x_i} \ket{x_1,0,i,0,0}
\end{aligned}
\end{equation}

As we see, if we apply $A$ and $A\otimes A$ at the same time, we obtain a diagonal operator that modifies a different part of the encoding, changing the amplitudes. For applying $B$, we need to use an ancilla to identify the elements $g(x_1,x_1,i,j)$ that, using our encoding, correspond to $c_v(x_1,0,i,j)$. As $\ket{a}= \ket{1}$ for the ancilla, both in the element encoding $\tau=1$ that is changed to $\tau=0$ and $\tau=0$, the ancilla qubit can be fully recovered, and the total operator combining all these transformations can be done. However, the total operator is not unitary; therefore, a Linear Combination of Unitaries (LCU) is used. The total operator is therefore 

\[
\begin{bmatrix}
A & B & 0 \\
0 & A\otimes A& 0 \\
0 & 0 & A\otimes A
\end{bmatrix}
\begin{bmatrix}
f  \\
g(x_1,0)   \\
g(x_1,x_2 \ne 0)
\end{bmatrix}
=
\begin{bmatrix}
Af + Bg(x_1,0) \\
A\otimes A g(x_1,0) \\
A\otimes A g(x_1,x_2 \ne 0)
\end{bmatrix}
\]

where we used $f$ for $\tau=0$ and $g$ for $\tau=1$ and we distinguish between the elements encoding diagonals (that contribute to $f$) and the other terms.

To sum up, we have proved that the encoding used represents the diagonal terms with a zero in the second dimension and that the shift used is equivalent to the desired change in the original encoding. Additionally, we have proved that the ancilla qubit can be recovered and that the operation can be done locally. The algorithm applies to any arbitrary number of time steps and lattice dimensions. Finally, we suggested that the total operator can be encoded by using an LCU procedure, making the whole operation unitary. The usage of LCU comes with the limitation of an exponentially lower probability of obtaining the right outcome after an arbitrary number of time steps. This limitation is not due to the encoding of the algorithm itself, but the non-unitarity of the collision operator.

\bibliography{apssamp}% Produces the bibliography via BibTeX.

\end{document}